\documentclass[preprint,12pt,3p,fleqn]{elsarticle}

\usepackage{amssymb}
\pdfoutput=1
\usepackage{amsthm}
\usepackage{amsmath}
\usepackage{multirow}
\usepackage{booktabs}
\usepackage{graphicx}
\usepackage{float}
\usepackage{url}

\usepackage{epstopdf}
\usepackage{pdflscape}

\usepackage{amsthm}
\usepackage{amsmath}
\usepackage[english]{babel}

\usepackage{subfigure}
\usepackage{threeparttable}

\usepackage{algorithm}
\usepackage{algorithmic}

\usepackage{xcolor}
\usepackage{color}

\usepackage{tabularx}
\usepackage{graphicx}
\usepackage{caption}
\usepackage{threeparttable}
\usepackage[figuresright]{rotating}
\usepackage[toc,page,title,titletoc,header]{appendix}

\makeatletter

\newcommand{\Rmnum}[1]{\expandafter\@slowromancap\romannumeral #1@}
\makeatother

\journal{Computer Methods in Applied Mechanics and Engineering}

\begin{document}
\captionsetup{font={scriptsize}}
\captionsetup[figure]{labelformat={default},labelsep=period,name={Fig.}}

\begin{frontmatter}
\title{A Block-based Adaptive Particle Refinement SPH Method for Fluid-Structure Interaction Problems}

\author[a]{Tianrun Gao}
\author[a]{Huihe Qiu}
\author[a,b,c]{Lin Fu\corref{cor1}}
\ead{linfu@ust.hk}
\cortext[cor1]{Corresponding author.}

\address[a]{Department of Mechanical and Aerospace Engineering, The Hong Kong University of Science and Technology, Clear Water Bay, Kowloon, Hong Kong}
\address[b]{Department of Mathematics, The Hong Kong University of Science and Technology, Clear Water Bay, Kowloon, Hong Kong}
\address[c]{Shenzhen Research Institute, The Hong Kong University of Science and Technology, Shenzhen, China}

\begin{abstract}

The multi-resolution method, e.g., the Adaptive Particle Refinement (APR) method, has been developed to increase the local particle resolution and therefore the solution quality within a  {pre-defined refinement zone} instead of using a globally uniform resolution for Smoothed Particle Hydrodynamics (SPH). However, sometimes, the targeted zone of interest can be varying, and the corresponding topology is very complex, thus the conventional APR method is not able to track these characteristics adaptively. In this study, a novel Block-based Adaptive Particle Refinement (BAPR) method is developed, which is able to provide the necessary local refinement flexibly for any targeted characteristic, and track it adaptively.  {In BAPR, the so-called activation status of the block array defines the refinement regions, where the transition and activated zones are determined accordingly. A regularization method for the generated particles in the newly activated blocks is developed to render an isotropic distribution of these new particles. The proposed method has been deployed for simulating Fluid-Structure Interaction (FSI) problems. A set of 2D FSI cases have been simulated with the proposed BAPR method, and the performance of the BAPR method is quantified and validated comprehensively.} In a word, the BAPR method is viable and potential for complex multi-resolution FSI simulations by tracking any targeted characteristic of interest.

\end{abstract}


\end{frontmatter}

\section{Introduction}
\label{section:1}

SPH is a meshless Lagrangian method, which can simulate many complex problems such as free surface flows, multi-phase flows, FSI, etc. In ocean engineering, FSI is a very important research problem for off-shore structures like wind turbines, tidal turbines and floating bodies. Traditional methods for FSI problems mainly resort to the Euler or Arbitrary-Lagrangian-Eulerian (ALE) formulations \cite{takashi1994ale}, where the mesh is adapted to track the structure surface. It will become more complicated when the free surface exists in the computational domain, for which the fluid interface tracking method must also be incorporated, such as the Volume-of-Fluid (VOF) or the level-set method. In terms of the coupling method between the fluid and the structure, the weak coupling and strong coupling strategies are usually adopted. However, even the weak coupling in the ALE system will greatly increase the complexity when the free surface is violent and the structure deforms intensively. As an alternative, the Immersed Boundary Method (IBM) \cite{peskin2002immersed} deploys the Eulerian mesh in the fluid field and the Lagrangian mesh in the solid field, whereas the data communications between two different mesh resolutions may result in the accuracy degeneration problem. 

Over the past decades, the meshless particle methods based on the Lagrangian concept for FSI problems have been extensively explored, e.g., the SPH methods \cite{sun2019study}\cite{zhang2021multi}\cite{o2021fluid}\cite{liu2013numerical} {\cite{khayyer2021coupled}\cite{zhan2019stabilized}\cite{shimizu2022sph}}, the Moving Particle Semi-implicit (MPS) method \cite{khayyer2019multi} {\cite{khayyer20213d}\cite{zha2021improved}}, the Finite Particle Method (FPM) \cite{zhang2019predicting}\cite{idelsohn2008unified}, and a set of coupling techniques with those methods \cite{zhang2018mps}\cite{fourey2017efficient}.  {Apart from the total Lagrangian concept, a set of ALE based SPH methods \cite{adami2013transport}\cite{sun2019consistent}\cite{jacob2021arbitrary}\cite{antuono2021delta} have accordingly been devised to make particle distribution more isotropic.} Studies using SPH method on FSI problems have been reported with various cases, e.g., water impact on the elastic beam  \cite{sun2019study}, flow induced flapping beam \cite{zhang2021multi}, dam breaking \cite{liu2013numerical}, sloshing \cite{chen2013investigation}, and two-phase problems \cite{sun2019study}. The SPH method successfully demonstrates its potential in simulating the complex fluids and its interaction with large-deformation structures,  {readers are referred to \cite{liu2019smoothed}\cite{gotoh2021entirely} for a comprehensive literature review.} 

In practice, the attention is usually focused on a small zone instead of the whole computational domain, thus the multi-resolution method with the local particle refinement is necessary to increase the local simulation accuracy while saving the overall computational cost. Till now, various multi-resolution strategies have been developed, such as the particle splitting technique \cite{feldman2007dynamic} to refine the coarse particles and the merging technique \cite{vacondio2013variable} to restore the coarse resolution beyond the interested zone.  {There is a set of studies deploying different resolutions for fluids and solids \cite{khayyer2021multi}\cite{ng2022improved}, where the interactions of particles between different resolutions are required.} Recently, the APR method \cite{barcarolo2014adaptive}\cite{chiron2018analysis} has been proposed by combining the particle splitting and deletion, and guard particles are deployed in the transition zones between different resolutions. This method is effective and efficient to investigate the solution details in the area of interest in the FSI problems, especially in violent impact flows.  {There are various studies \cite{sun2019extension}\cite{hermange20193d} on FSI problems using APR method, where this multi-resolution method has shown a great potential.}

The APR method is generally suitable for problems with the refinement zones, which are known as a priori. However, in most cases, the area of interest is unknown before the simulation, and the targeted refinement area is varying in both space and time, rendering that the APR zones cannot be determined before simulation. Yang et al. \cite{yang2021smoothed} propose an adaptive refinement method to enhance the local resolution on the fluid surface or the fluid-solid interface. However, this method can only be used to enhance the accuracy at the geometry boundaries, with the need of frequent particle merging and direct particle interaction across different resolutions, which may lead to the accuracy degeneration. 

In the Eulerian Cartesian mesh methods, the Adaptive Mesh Refinement (AMR) method is usually employed with certain criteria for capturing the phase boundaries, the shock waves, or the vortex clustering regions \cite{gunney2013scalable}\cite{fu2017single}\cite{fu2017novel}. For example, the shock wave is supposed to be identified and captured with the refined resolution for compressible flows, and for incompressible flows, the fluid characteristics, e.g., the vortex or the pressure gradient, may be the targeted quantity for refinement. For FSI problems, the flow surrounding the structure is prone to generating vortex, and the vortex may affect the dynamic behavior of the structure, therefore the vorticity may be the targeted criterion. There have been a number of studies with block-based multi-resolution mesh methods to investigate the FSI problems. Liu and Hu \cite{liu2018block} combine the IBM method and the block-based AMR method to study the vortex induced vibration problem in the incompressible flows. Deiterding and Wood \cite{deiterding2013parallel} employ the block-structured AMR method to investigate the explosion FSI problems in compressible flows. Though the APR method is similar to AMR in many aspects, and has been used in the violent flow problems and the FSI problems, the APR method cannot track the targeted characteristic adaptively. To the best of authors' knowledge, there are few reports on the FSI problems with the fully adaptive particle refinement strategy.

Inspired by the block-based AMR method in Eulerian framework \cite{fu2017single}, in this work, we will develop a block-based adaptive particle refinement (BAPR) SPH method for FSI problems.  {The difference is that the refinement scale of the block in BAPR is larger than that of the mesh in the AMR method.}
The remainder of this paper is organized as follows. In section~\ref{section:2}, the basic SPH formulations for fluid and solid governing equations will be presented. In section~\ref{section:3}, the BAPR algorithms for SPH will be elaborated in detail, including the block array assignment, the characteristic identification, the transition zone identification algorithm, and the regularization strategy for the newly generated particles. In section~\ref{section:4}, a set of benchmark cases will be presented and analyzed, including water impact on an elastic beam, dam-breaking through an elastic beam, flapping beam induced by the flow over a cylinder, flow over an inclined elliptical cylinder, and body entry problems. The performance of the proposed BAPR method will be analyzed and validated. In section~\ref{section:5}, concluding remarks and discussions will be given.

\section{Numerical methods for FSI problems}
\label{section:2}

\subsection{Governing equations for the fluid field }
\label{subsection:GF}

The Navier–Stokes equations for fluids can be written in the Lagrangian form as 
\begin{equation}
\label{eq:sph_comp}
\left\{
\begin{array}{l}
\begin{aligned}
\vspace{1ex}
&\dfrac{D\rho}{Dt} = - \rho \boldsymbol{\nabla} \cdot \boldsymbol{v},\vspace{1ex} \\
&\dfrac{D\boldsymbol{v}}{Dt} = - \dfrac{\boldsymbol{\nabla} p}{\rho} + \dfrac{\eta }{\rho} \boldsymbol{\nabla} ^{2} \boldsymbol{v} +\boldsymbol{f},  \\
&\dfrac{D\boldsymbol{r}}{Dt} = \boldsymbol{v},
\end{aligned}
\end{array}
\right.
\end{equation}
where $\rho$, $\boldsymbol{v}$, $p$, $\eta$, $\boldsymbol{f}$ and $\boldsymbol{r}$ denote the fluid density, velocity, pressure, dynamic viscosity, body force and the particle coordinate, respectively. The Equation-of-State (EOS) is used to relate the pressure and the density, which is usually written as 
\begin{equation}
\label{eq:eos}
p=c_{0}^{2} \left( \rho - \rho_{0} \right),
\end{equation}
 {where $c_{0}$ denotes the artificial sound speed and is constrained by $c_{0} \geq 10 \left( U_{max}, \sqrt{\frac{p_{max}}{\rho_0}} \right)$ \cite{sun2017deltaplus}\cite{hammani2020detailed} to ensure the weak-compressibility condition, and $U_{max}$ and $p_{max}$ are the expected maximum velocity and pressure, respectively.}

The SPH-discretized approximation for equation (\ref{eq:sph_comp}) can be given as 
\begin{equation}
\label{eq:discr_sph}
\left\{
\begin{array}{l}
\begin{aligned}
\vspace{1ex}
\dfrac{D {\rho}_{i}}{D t} &={\rho}_{i} \sum_{j}  \boldsymbol{v}_{i j} \cdot \boldsymbol{\nabla}_{i} W_{i j} V_{j}+\delta h c_{0} \sum_{j} \boldsymbol{\Phi}_{i j} \cdot \boldsymbol{\nabla}_{i} W_{i j} V_{j},\\
\dfrac{D \boldsymbol{v}_{i}}{D t} &=  -\dfrac{1}{\rho_i} \sum_{j} \left(p_{i}+p_{j}\right) \boldsymbol{\nabla}_{i} W_{i j} V_{j} +
\dfrac{1}{\rho_i}\sum_{j} \frac{\left(\eta_{i}+\eta_{j}\right) \boldsymbol{r}_{i j} \cdot \boldsymbol{\nabla}_{i} W_{i j}V_{j}}{\left(r_{i j}^{2}+ 0.01h^{2}_{i}   \right)} \boldsymbol{v}_{i j} \\
&+\alpha h c_{0} \sum_{j} \pi_{ij} \boldsymbol{\nabla}_{i} W_{i j} V_{j} + \boldsymbol{f},\\
\dfrac{D\boldsymbol{r_{i}}}{Dt} &= \boldsymbol{v}_{i},
\end{aligned}
\end{array}
\right.
\end{equation}
where $\boldsymbol{v}_{i j}=\boldsymbol{v}_{i} - \boldsymbol{v}_{j}$, $\boldsymbol{r}_{i j}=\boldsymbol{r}_{i} - \boldsymbol{r}_{j}$, and $V_{j}=m_{j}/\rho_{j}$ is the volume of the particle. $\boldsymbol{\nabla}_{i} W_{i j}$ denotes the gradient of the kernel function $W(\left| \boldsymbol{r}_{ij}\right|,h)$. In this work, the Gaussian kernel is adopted with a smoothing length of $h=1.2\Delta {x}$, where $\Delta {x}$ is the initial particle spacing, and the cut-off parameter is set as 3. The artificial viscosity \cite{monaghan1992smoothed} is deployed in the momentum equation to stabilize the computation for inviscid flows, with the term expressed as
\begin{equation}
\label{eq:delta_term}
{\pi}_{i j} =  \left( \boldsymbol{v}_{j}-\boldsymbol{v}_{i} \right) \cdot \dfrac{ \boldsymbol{r}_j - \boldsymbol{r}_i }{\left| { \boldsymbol{r}_j - \boldsymbol{r}_i}\right| ^{2} } .
\end{equation}
In order to alleviate the pressure fluctuation in the computational field,  {a simplified version of the dissipation term \cite{molteni2009simple},} i.e.,
\begin{equation}
\boldsymbol{\Phi}_{i j} = 2\left( \rho_{j}- \rho_{i} \right) \dfrac{ \boldsymbol{r}_j - \boldsymbol{r}_i }{\left| { \boldsymbol{r}_j - \boldsymbol{r}_i}\right| ^{2} },
\end{equation}
is introduced in the continuity equation. The two coefficients are chosen as $ \delta=0.1 $  {\cite{molteni2009simple}\cite{marrone2011delta}} and $ \alpha =0.02 $ \cite{marrone2011delta} in this paper. 

In practical simulations, the ALE-based SPH formulation is usually employed to avoid the numerical voids which typically appear with the total Lagrangian formulation \cite{adami2013transport}\cite{oger2016sph}\cite{sun2019consistent}\cite{antuono2021delta}, and a transport velocity $\tilde{\boldsymbol{v}}$ which is generated by the driving force due to the irregular particle distribution is incorporated in the ALE framework to regularize the particle distribution. The ALE formulation is written as
\begin{equation}
\left\{
\begin{array}{l}
\begin{aligned}
\vspace{1ex}
&\dfrac{d\rho}{dt} =  {- \rho \boldsymbol{\nabla} \cdot \tilde {\boldsymbol{v}} -\boldsymbol{\nabla} \cdot[\rho(\boldsymbol{v}-\tilde{\boldsymbol{v}})]}, \\
&\dfrac{d\boldsymbol{v}}{dt} = - \dfrac{\boldsymbol{\nabla} p}{\rho} -\boldsymbol{\nabla} \cdot[\boldsymbol{v} \otimes (\boldsymbol{v}-\tilde{\boldsymbol{v}})]
+\boldsymbol{v} \boldsymbol{\nabla} \cdot(\boldsymbol{v}-\tilde{\boldsymbol{v}})+\dfrac{\eta }{\rho} \boldsymbol{\nabla} ^{2} \boldsymbol{v} + \boldsymbol{f} ,  \\
&\dfrac{d\boldsymbol{r}}{dt} = \tilde{\boldsymbol{v}}.
\end{aligned}
\end{array}
\right.
\end{equation}
Here, the ALE derivative $\frac{d \phi }{dt}$ is used instead of the full derivative $\frac{D \phi }{Dt}$. The corresponding SPH-discretized formulations of the ALE form can be expressed as
\begin{equation}
\label{eq:discr_sph_tranv}
\left\{
\begin{array}{l}
\begin{aligned}
\vspace{1ex}
\dfrac{d {\rho}_{i}}{d t} &= {\rho}_{i} \sum_{j}  \tilde{\boldsymbol{v}}_{i j} \cdot \boldsymbol{\nabla}_{i} W_{i j} V_{j}+\delta h c_{0} \sum_{j} \boldsymbol{\Phi}_{i j} \cdot \boldsymbol{\nabla}_{i} W_{i j} V_{j} \\
&+ \sum_{j} [\rho_{i} \delta \boldsymbol{v}_{i} + \rho_{j} \delta \boldsymbol{v}_{j}]\cdot \boldsymbol{\nabla}_{i} W_{i j} V_{j},\\

\dfrac{d \boldsymbol{v}_{i}}{d t} &=  -\dfrac{1}{\rho_i} \sum_{j} \left(p_{i}+p_{j}\right) \boldsymbol{\nabla}_{i} W_{i j} V_{j} +\alpha h c_{0} \sum_{j} \pi_{ij} \boldsymbol{\nabla}_{i} W_{i j} V_{j} \\
&+\sum_{j} [\boldsymbol{v}_{i} \otimes \delta \boldsymbol{v}_{i}+ \boldsymbol{v}_{j} \otimes \delta \boldsymbol{v}_{j}]\cdot \boldsymbol{\nabla}_{i} W_{i j} V_{j} - \boldsymbol{v}_{i} \sum_{j} [\delta \boldsymbol{v}_{j}- \delta \boldsymbol{v}_{i}]\cdot \boldsymbol{\nabla}_{i} W_{i j} V_{j} \\
&+\dfrac{1}{\rho_i}\sum_{j} \frac{\left(\eta_{i}+\eta_{j}\right) \boldsymbol{r}_{i j} \cdot \boldsymbol{\nabla}_{i} W_{i j}V_{j}}{\left(r_{i j}^{2}+ 0.01h^{2}_{i}   \right)} \boldsymbol{v}_{i j}+ \boldsymbol{f},\\

\dfrac{d\boldsymbol{r_{i}}}{dt} &= \tilde{\boldsymbol{v}}_{i},
\end{aligned}
\end{array}
\right.
\end{equation}
where  {$\delta \boldsymbol{v}=\tilde{\boldsymbol{v}}-\boldsymbol{v}$} is the shifting velocity.  {The shifting velocity is driven by a background pressure gradient induced by the irregular particle distribution  \cite{adami2013transport}. As pointed out in \cite{adami2013transport}, the choice of background pressure is arbitrary. Owing to this reason, we introduce a term $0.2\left(\frac{W_{i j}}{W(\Delta x,h)}\right)^{4}$ into the conventional background pressure gradient referring to \cite{monaghan2000sph} and \cite{antuono2021delta},  where the added term can prevent the clustering of particles effectively,} and then the background pressure gradient $\boldsymbol{\nabla} P_{ {B}}$ is evaluated by
\begin{equation}
\label{eq:background_P}
\begin{aligned}
\boldsymbol{\nabla} P_{ {B}} = -\rho_{0} c_{0}^{2}  \sum_{j} \left[1+0.2\left(\dfrac{W_{i j}}{ {W(\Delta x,h)}}\right)^{4}  \right] \boldsymbol{\nabla}_{i} W_{i j} V_{j}.
\end{aligned}
\end{equation}

 {It is known that the particle shifting or transport velocity may lead to the violation of the volume conservation \cite{lyu2022further}\cite{krimi2020wcsph}\cite{jandaghian2022stability}, the main reason is due to the error accumulation of the particle shifting for the particles in the surface area.}  {Considering this, the background pressure gradient Eq. (\ref{eq:background_P}) will only be used for the particles in the non-surface area. For the particles in the surface area, Eq. (\ref{eq:background_P}) is turned off, which means that their shifting velocity is set as zero. Herein, a free surface detection method \cite{marrone2010fast} is employed to identify the free surface particles, and the `surface area' mentioned here means the region within the radius of $4\Delta x$ of the detected free surface particles.}  {Furthermore, in the scenarios with violent flows, the shifting velocity $\delta \boldsymbol{v}$ is constrained to be within 0.05 times of the norm of the real velocity $ \left| \boldsymbol{v} \right| $, written as
\begin{equation}
 \left| \delta \boldsymbol{v} \right| < 0.05\left|\boldsymbol{v} \right| .
 \label{eq:dv_con}
\end{equation}
For the cases without violent flows, this constraint is not used. }To achieve this goal, a free surface detection method is employed to identify the free surface particles \cite{marrone2010fast}. For boundary particles, the transport velocity will be set as zero.

For the time integration, the established Kick-Drift-Kick (KDK) \cite{monaghan2005smoothed}\cite{adami2013transport}\cite{zhang2017generalized} scheme is employed, i.e.,
\begin{equation}
	\label{eq:scheme}
	\begin{array}{l}
		\begin{aligned}
			&\boldsymbol{v}^{n+\frac{1}{2}} = \boldsymbol{v}^{n} + \frac{\Delta t_{F}}{2} {\left( \dfrac{d \boldsymbol{v}}{d t} \right)}^{n},\\
			&\boldsymbol{\tilde{v}}^{n+\frac{1}{2}} = \boldsymbol{v}^{n+\frac{1}{2}} + \frac{\Delta t_{F}}{2 {\rho_{0}}} \boldsymbol{\nabla} P_{ {B}}^{n},\\
			&\boldsymbol{x}^{n+1} = \boldsymbol{x}^{n} + {\Delta t_{F}} \boldsymbol{\tilde{v}}^{n+\frac{1}{2}},\\
			&\rho^{n+1} = \rho^{n} + \Delta t_{F} {\left(  \dfrac{d\rho}{d t} \right)}^{n+\frac{1}{2}},\\
			&\boldsymbol{v}^{n+1} = \boldsymbol{v}^{n+\frac{1}{2}} + \frac{\Delta t_{F}}{2} {\left( \dfrac{d \boldsymbol{v}}{d t} \right)}^{n+1}.\\
		\end{aligned}
	\end{array}
\end{equation}
 {In this scheme, the particle field values are calculated in sequence in Eq. (\ref{eq:scheme})}. The time step $\Delta t_{F}$ for the fluid evolution is defined as 
\begin{equation}
\begin{aligned}
& {\Delta t_{a} = 0.25 \sqrt{\dfrac{h}{\left| \boldsymbol{a} \right|_{max}}}, \text{ } \Delta t_{v} = 0.25 \dfrac{h}{c_{0}+{\left| \boldsymbol{v} \right|_{max}}}},\\
& {\Delta t_{F} = min(\Delta t_{a}, \Delta t_{v}).}
\end{aligned}
\end{equation}
\subsection{Governing equations for the solid field }

The Lagrangian form of the governing equations for the solid dynamics can be expressed as 
\begin{equation}
\dfrac{D\boldsymbol{v}}{Dt} = - \dfrac{1}{\rho_{S}} \boldsymbol{\nabla}_{0} \cdot \boldsymbol{\rm P} + \boldsymbol{f},
\end{equation}
where the operator $\boldsymbol{\nabla}_{0}$ is in regard with the static initial coordinates, and ${\boldsymbol{\rm P}}$ denotes the first Piola–Kirchhoff stress tensor. Assuming that the density of the solid $ \rho_{S} $ is invariant, the discretized form with the SPH approximation can be written as
\begin{equation}
\label{eq:solid}
\dfrac{D \boldsymbol{v}_{a}}{D t} =  -\dfrac{1}{\rho_{S}} \sum_{b} \left(\boldsymbol{\rm P}_{a} \boldsymbol{\rm L}_{0a}^{-1} + \boldsymbol{\rm P}_{b} \boldsymbol{\rm L}_{0b}^{-1} \right) \boldsymbol{\nabla}_{0a} W_{0ab} V_{b} + \boldsymbol{f},
\end{equation}
 {where the particle `$b$' denotes the neighbor of particle `$a$'}. Here the correction matrix \cite{randles1996smoothed} is given as 
\begin{equation}
\boldsymbol{\rm L}_{0a} = \sum_{b} ( \boldsymbol{ x}_{0b} -\boldsymbol{ x}_{0a} ) \otimes \boldsymbol{\nabla}_{0a} W_{0ab} V_{b},
\end{equation}
and $\boldsymbol{ x}_{0}$ is the initial position of the particle. In this work, the Saint Venant-Kirchhoff constitutive model for the structure is used. The stress tensor in Eq. (\ref{eq:solid}) is related to the deformation in the solid field. Specifically,  the displacement gradient $\boldsymbol{\rm F}$ can be expressed as 
\begin{equation}
\boldsymbol{\rm F} = \dfrac{d \boldsymbol{x} }{d \boldsymbol{x}_{0}},
\end{equation}
and the corresponding discretized form with the SPH approximation is written as 
\begin{equation}
\boldsymbol{\rm F}_{a} = \sum_{b} ( \boldsymbol{x}_{ b} -\boldsymbol{ x}_{a} ) \otimes \boldsymbol{\rm L}_{0a}^{-1} \boldsymbol{\nabla}_{0a} W_{0ab} V_{b}.
\end{equation}
Then, the deformation tensor Green–Lagrange strain can be derived using
\begin{equation}
\boldsymbol{\rm E} = \dfrac{1}{2}( { {\rm \boldsymbol{\rm F}}_{a}^{\rm T}} { \boldsymbol{\rm F}_{a}} - \boldsymbol{\rm I} ),
\end{equation}
and the second Piola–Kirchhoff stress tensor $\boldsymbol{\rm S}$ is obtained through
\begin{equation}
\boldsymbol{\rm S} = \lambda tr(\boldsymbol{\rm E} ) \boldsymbol{\rm I} + 2 \mu \boldsymbol{\rm E},
\end{equation}
where $\lambda$ and $\mu$ are Lam\'e parameters. At last, the first Piola–Kirchhoff stress tensor is obtained with
\begin{equation}
\boldsymbol{\rm P} = \boldsymbol{\rm F} \boldsymbol{\rm S}.
\end{equation}

The Kick-Drift-Kick (KDK) \cite{monaghan2005smoothed}\cite{adami2013transport}\cite{zhang2017generalized} scheme is employed for the time integration, i.e.,
\begin{equation}
	\label{eq:scheme1}
	\begin{array}{l}
		\begin{aligned}
			&\boldsymbol{v}^{n+\frac{1}{2}} = \boldsymbol{v}^{n} + \frac{\Delta t_{S}}{2} {\left( \dfrac{D \boldsymbol{v}}{D t} \right)}^{n},\\
			&\boldsymbol{x}^{n+1} = \boldsymbol{x}^{n} + {\Delta t_{S}} \boldsymbol{v}^{n+\frac{1}{2}},\\
			&\boldsymbol{v}^{n+1} = \boldsymbol{v}^{n+\frac{1}{2}} + \frac{\Delta t_{S}}{2} {\left( \dfrac{D \boldsymbol{v}}{D t} \right)}^{n+1},\\
		\end{aligned}
	\end{array}
\end{equation}
with which there is no transport velocity. The time step $\Delta t_{S}$ for the solid is chosen as 
\begin{equation}
\Delta t_{S} = 0.5h/c_{s},\text{ }c_{s}= \sqrt{\frac{E (1-\nu)}{\rho_{S}(1+\nu)(1-2\nu)}},
\end{equation}
where $c_{s}$, $E$ and $\nu$ denote the elastic wave speed, Young's modulus and Poisson's ratio, respectively.
\subsection{Coupling algorithms between fluid and solid}

For FSI problems, the timestep of the fluid phase is typically much larger than that of the solid phase, thus within one fluid timestep, the solid field may be advanced with many timesteps. Consequently, a weak coupling strategy is usually employed, where the motions of fluid particles and solid particles are solved separately. More specifically, at the start of each fluid timestep, with the solid particle fixed, the fluid particles will advance with one fluid timestep. Then the updated values of fluid particles are fixed, the solid particles will advance with many solid timesteps within one fluid timestep. More details are referred to \cite{sun2021accurate}.

For the fluid particles, the solid particles provide the necessary boundary conditions. The fixed boundary particle method is employed and the fluid variables of the boundary particles are determined by the technique proposed by Adami et al. \cite{adami2012generalized}. The momentum equation for the evolution of the fluid particles in the ALE form is given as
\begin{equation}
\label{eq:discr_couple_f}
\begin{array}{l}
\begin{aligned}

\dfrac{d \boldsymbol{v}_{i}}{d t} &=  -\dfrac{1}{\rho_{i}} \sum_{j\in F} \left(p_{i}+p_{j}\right) \boldsymbol{\nabla}_{i} W_{i j} V_{j}  + \dfrac{1}{\rho_i}\sum_{j\in F} \frac{\left(\eta_{i}+\eta_{j}\right) \boldsymbol{r}_{i j} \cdot \boldsymbol{\nabla}_{i} W_{i j}V_{j}}{\left(r_{i j}^{2}+ 0.01h^{2}_{i}   \right)} \boldsymbol{v}_{i j} +\alpha h c_{0} \sum_{j\in F} \pi_{ij} \boldsymbol{\nabla}_{i} W_{i j} V_{j}\\
&+\sum_{j } [\boldsymbol{v}_{i} \otimes \delta \boldsymbol{v}_{i}+ \boldsymbol{v}_{j} \otimes \delta \boldsymbol{v}_{j}]\cdot \boldsymbol{\nabla}_{i} W_{i j} V_{j} - \boldsymbol{v}_{i} \sum_{j} [\delta \boldsymbol{v}_{j}- \delta \boldsymbol{v}_{i}]\cdot \boldsymbol{\nabla}_{i} W_{i j} V_{j}\\
&-\dfrac{1}{\rho_i} \sum_{j\in S | B} \left(p_{i}+p_{j}^{\prime}\right) \boldsymbol{\nabla}_{i} W_{i j} V_{j} + \dfrac{1}{\rho_i}\sum_{j\in S | B} \frac{\left(\eta_{i}+\eta_{j}^{\prime}\right) \boldsymbol{r}_{i j} \cdot \boldsymbol{\nabla}_{i} W_{i j}V_{j}}{\left(r_{i j}^{2}+ 0.01h^{2}_{i}   \right)} \boldsymbol{v}_{i j}^{\prime}\\
&+\alpha h c_{0} \sum_{j\in S|B} \pi_{ij}^{\prime} \boldsymbol{\nabla}_{i} W_{i j} V_{j} + \boldsymbol{f},

\end{aligned}
\end{array}
\end{equation}
where the variables with the superscript $'$, e.g., $p_{j}^{\prime}$, denote the virtual fluid variables of solid or boundary particles. $F$ and $S|B$ represent the fluid domain and the solid or boundary domain, respectively. 

For the solid particles, the fluid particles provide external force loading. The momentum equation for the evolution of the solid particles in the total Lagrangian form is given as
\begin{equation}
\label{eq:discr_couple_s}
\begin{array}{l}
\begin{aligned}

\dfrac{D \boldsymbol{v}_{a}}{D t} &= -\dfrac{1}{\rho_{S}} \sum_{b \in S} \left(\boldsymbol{\rm P}_{a} \boldsymbol{\rm L}_{0a}^{-1} + \boldsymbol{\rm P}_{b} \boldsymbol{\rm L}_{0b}^{-1} \right) \boldsymbol{\nabla}_{0a} W_{0ab} V_{b}\\
&+\dfrac{1}{\rho_{S}} \sum_{b\in F} \left(p_{a}^{\prime}+p_{b}\right) \boldsymbol{\nabla}_{a} W_{ab} V_{b} +
\dfrac{1}{\rho_S}\sum_{b\in F} \frac{\left(\eta_{a}^{\prime}+\eta_{b}\right) \boldsymbol{r}_{ab} \cdot \boldsymbol{\nabla}_{a} W_{ab}V_{b}}{\left(r_{ab}^{2}+ 0.01h^{2}_{a}   \right)} \boldsymbol{v}_{ab}^{\prime}\\
&+\alpha h c_{0} \dfrac{\rho^{\prime}_{a}}{\rho_{S}} \sum_{b\in F} \pi_{ab}^{\prime} \boldsymbol{\nabla}_{a} W_{ab} V_{b} + \boldsymbol{f}.

\end{aligned}
\end{array}
\end{equation}

\section{The improved APR method and the new BAPR method}
\label{section:3}
\subsection{Improved APR method}
\begin{figure}
\centering
\includegraphics[width=0.7\textwidth]{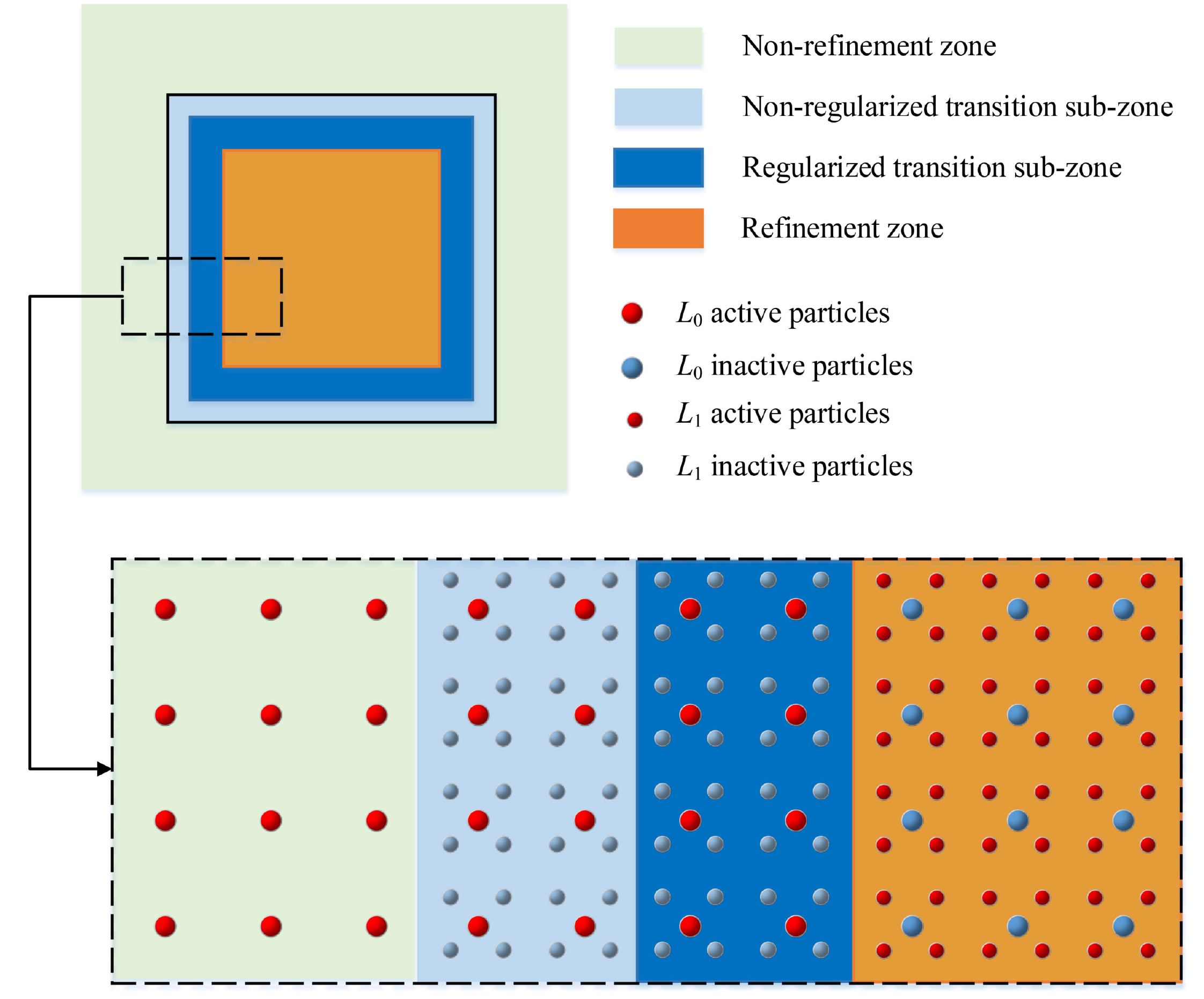}
\caption{Schematic of the APR method: the refinement zone is constrained in a  {pre-defined refinement zone}, where the edge region is the transition zone and the interior region is the refinement zone (orange), and the outer layer of the transition zone is the non-regularized transition sub-zone (light blue) and the inner layer is the regularized transition sub-zone (dark blue). }
\label{APR}
\end{figure}

A multi-resolution method of so-called APR is proposed initially by Barcarolo et al. \cite{barcarolo2014adaptive} and improved by Chiron et al. \cite{chiron2018analysis}. This method aims to increase the local field accuracy in the  {pre-defined refinement zone} with the refined particles while preserving the coarse resolution in the left area. The particles of different resolutions are defined as the different sets of particles, indicated with  $ L_k $, where $ k $ is the level of the particles. If there are two levels of particles, $ L_0 $ indicates the coarse particles and $ L_1 $ the refined particles, as shown in Fig. \ref{APR}. In this method, the particles of different levels are marked as `active' and `inactive', i.e., if the particle is `active', it will take part in the physical time evolution by solving the governing equations; if the particle is `inactive', it will obtain its field values such as density and velocity by the so-called Shepard interpolation method  {\cite{shepard1968two}}. 

After the  {pre-defined refinement zone} is deployed, for the $ L_0 $ particles, once a $ L_0 $ particle enters the  {pre-defined refinement zone},  four new $ L_1 $ particles (for 2D problems) will be generated while the $ L_0 $ particle is retained and marked as `inactive'. If the  $ L_0 $ particle exits from the  {pre-defined refinement zone}, it will be marked as `active' again. For the  $ L_1 $ particles, once they exit from the  {pre-defined refinement zone}, they will be deleted. Notably, at the edge of the  {pre-defined refinement zone}, there is a transition zone, which provides the support for information exchange between the two levels. Thus, for all the `active' $ L_1 $ particles, their neighbor particles are all taken from the same level, since the information exchange between two levels is provided by the particles in the transition zone.

In this study, the standard APR method is improved by introducing a regularized transition zone. In fact, when the $ L_0 $ particles enter the transition zone, the distribution of $ L_1 $ particles generated from the $ L_0 $ particles will not be isotropic, which will affect the accuracy of the following SPH simulations.  {In order to ensure the simulation accuracy of $ L_1 $ `active' particles, we have to render those $ L_1 $ `inactive' particles, which are close to $ L_1 $ `active' particles, more isotropic}. To this end, the transition zone is further divided into two parts, i.e., the non-regularized transition sub-zone and the regularized transition sub-zone. In the regularized transition sub-zone, the transport velocity of $ L_1 $ particles driven by the background pressure gradient force Eq. (\ref{eq:background_P}) is used to regularize the particle distribution; for the non-regularized transition sub-zone, the shifting velocity of the particles is set to be zero considering their incomplete particle kernel support. In this way, the neighbors of all the `active' $ L_1 $ particles have an isotropic distribution due to the regularized transition sub-zone, which improves the accuracy of the simulations.  {For $ L_0 $ particles, all the `inactive' particles, except from those in the surface area, will be regularized using the background pressure gradient force Eq. (\ref{eq:background_P}).}
 {It is worth noting that the size of the refinement zone is flexible, and there are no extra requirements on the size of the APR zone. For the non-regularized transition sub-zone, the thickness is set as $ 4\Delta x_1 $ ($\Delta x_0$ and $\Delta x_1$ denote the initial particle spacing at the resolution level $0$ and $1$,  respectively), which ensures that all the particles with incomplete kernel support are involved. For the regularized transition sub-zone, a thickness of larger than $ 4\Delta x_1 $ is recommended. }

\subsection{Proposed BAPR method}

In the section above, the APR method is only applicable for cases with the  {pre-defined refinement zone}. However, for some dynamic flow features, of which the motion is unknown beforehand, the APR method will not function as expected. For this type of problem, a fully adaptive multi-resolution method should be established to track the targeted characteristics with the refined resolutions. To this end, inspired by the AMR method, a new BAPR method is proposed in this section. The new method can track any characteristic and refine the particles within the activated blocks adaptively. Firstly, the corresponding blocks for particle refinement are identified. Secondly, the transition and refinement zones in the identified blocks are defined to determine the activation status of all the particles. Finally, the regularization technique will be implemented to render an isotropic particle distribution for the newly generated refined particles. 

\subsubsection{Identification of the activated blocks}

\begin{figure}
\centering
\includegraphics[width=0.7\textwidth]{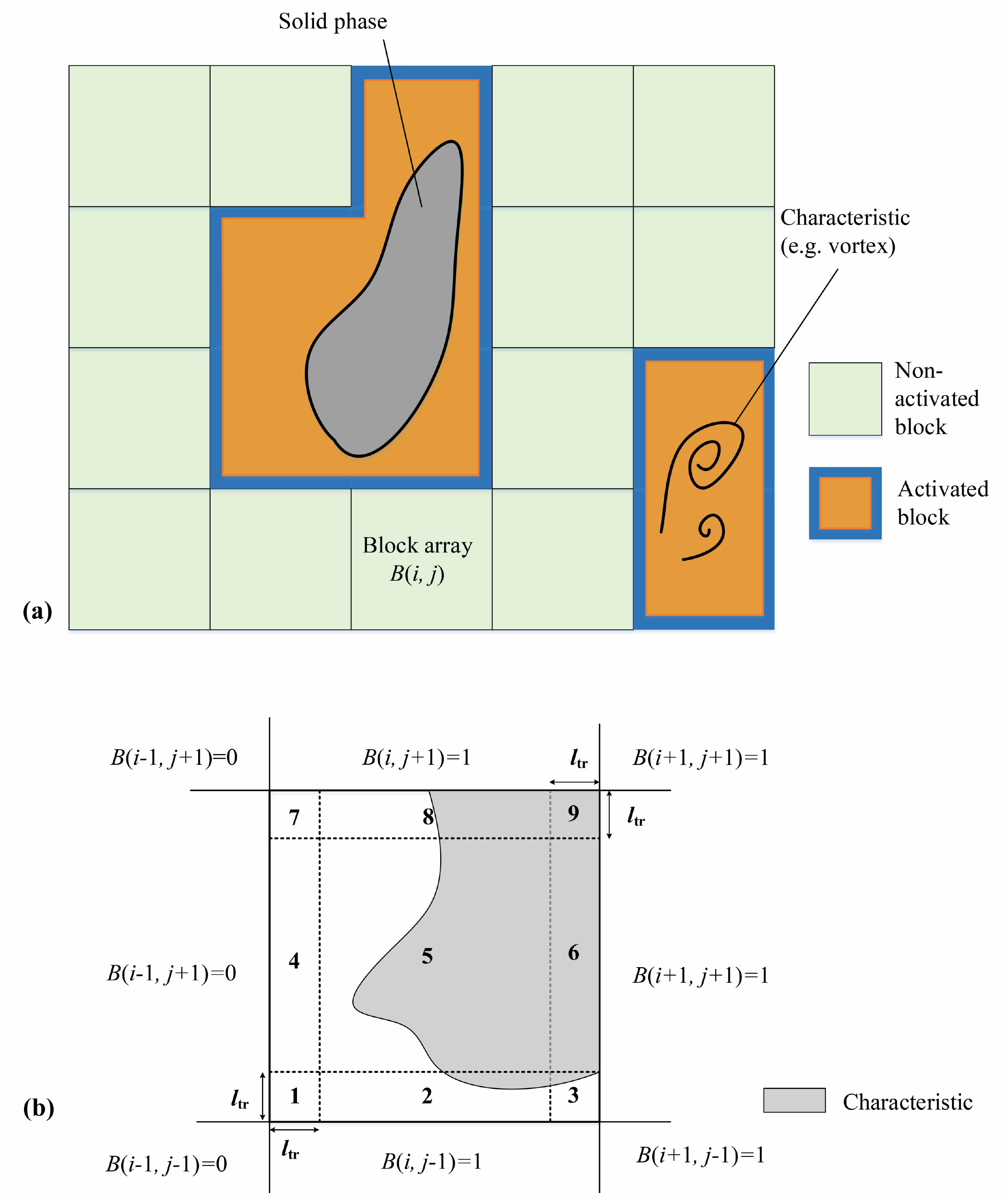}
\caption{Schematic of the proposed BAPR method: (a) the block regions around the solid phase or where involving the targeted characteristic are identified as the refinement zones, in which the regions with the orange color are the refinement zones and blue the transition zones;  {(b) block identification method: the block is divided into 9 subdomains (in this figure $l_{tr}$ is the thickness of the transition zone, which will be introduced with details in section \ref{sect:transition}). If a subdomain $S$ contains the targeted characteristic, the corresponding neighbor blocks of the subdomain $S$ will be activated.}}
\label{BAPR}
\end{figure}

The whole computational domain is partitioned into an array of blocks, where the blocks are the candidate refinement regions consisting of the refinement and transition zones.  {The choice of the block size in the BAPR method is quite flexible, and the block length $ L_x$ and height $ L_y$ are determined by the user. A suggestion is that the block size should not be too small, because the block should be at least larger than 2 times of the transition zone thickness to ensure that there are `active' particles in the refinement zone of this block, as shown in Fig. \ref{BAPR} and \ref{P_m}}. As shown in Fig. \ref{BAPR}(a), the light green zone represents the non-activated blocks, where only the $ L_0 $ particles exist. Once there are solid particles in one specific block, it will be identified as an activated block. Likewise, if some features, e.g., vortex, form and reach the prescribed threshold, the corresponding block will be identified as an activated block as well.  {Specifically, the block identification method is shown in Fig. \ref{BAPR}(b), where the block is divided into 9 subdomains. If the targeted characteristic exists in a subdomain $S$, the corresponding neighbor blocks of the subdomain $S$ will be activated. This operation ensures that the targeted characteristic is sufficiently refined throughout the simulation.}

\subsubsection{Identification of the transition and refinement zones}
\label{sect:transition}
\begin{figure}
\centering
\includegraphics[width=0.7\textwidth]{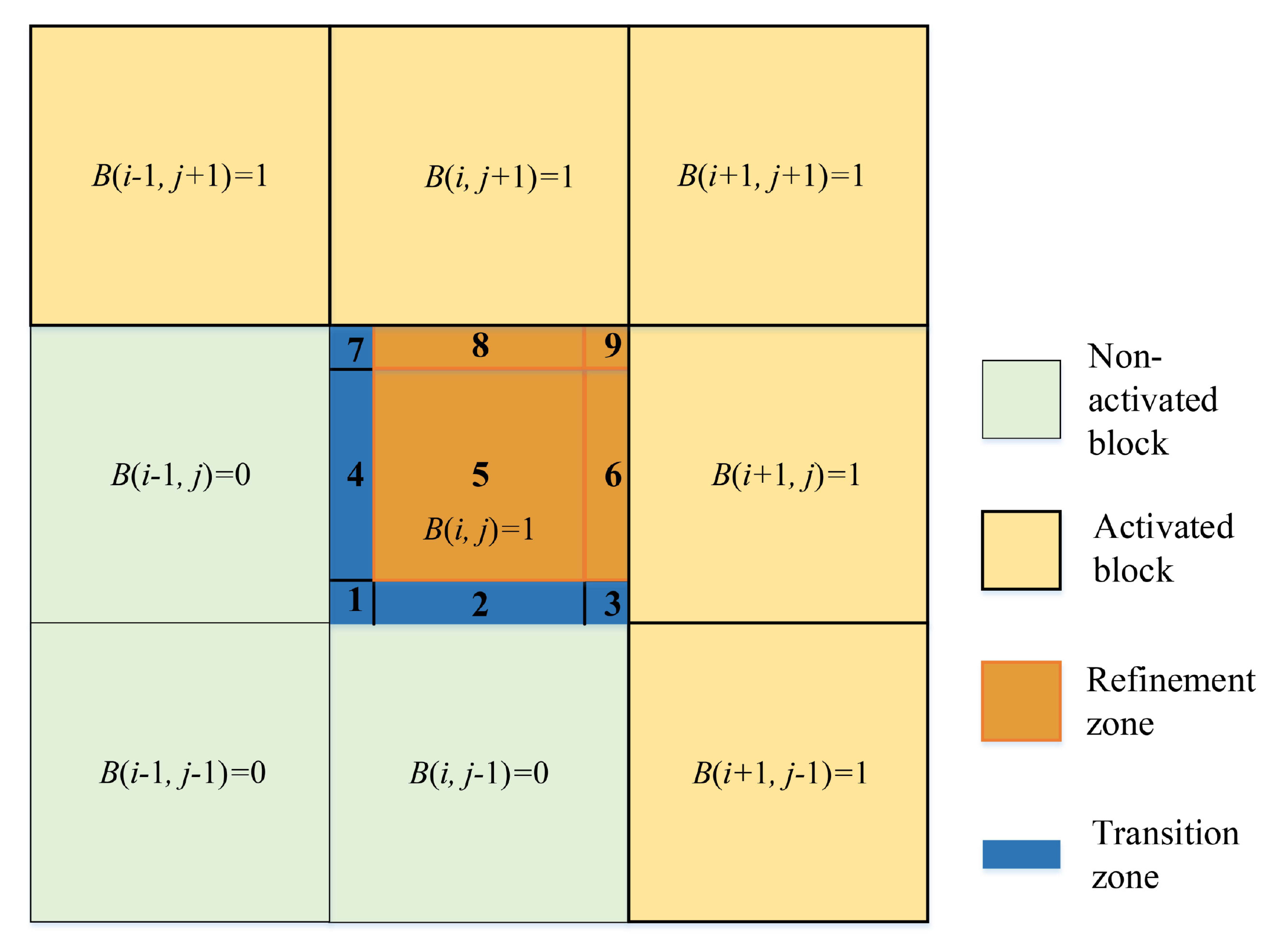}
\caption{Schematic of the transition zones for block $B(i,j)$: for subdomains $1\sim 9$ in the targeted block $B(i,j)$, if not all its neighbors are activated blocks, this subdomain will be identified as a transition zone.}
\label{P_m}
\end{figure}

The transition zone is important for the simulation accuracy of the refined particles. Different from the standard APR method, the identification of the transition zone in BAPR requires a new algorithm. For this purpose, one block is partitioned into nine subdomains, and marked with the index from 1$\sim $9, as shown in Fig. \ref{P_m}. Taking the block $B(i,j)$ for instance, each subdomain in this block will be determined as the transition zone or the refinement zone by examining the status of its neighbor blocks. If all the neighbor blocks of one subdomain are the activated blocks, this subdomain will be identified as the refinement zone; otherwise, it will be identified as the transition zone. 

Similarly, in order to ensure an isotropic distribution of the refined particle neighbors, the regularized transition sub-zone will be identified.  {The thickness of the transition zone $l_{tr}$ is $ 10\Delta x_1 $ in BAPR, and the thicknesses of the regularized transition sub-zone and the non-regularized transition sub-zone are $ 6\Delta x_1 $ and $ 4\Delta x_1 $, respectively. } For the particles in the regularized transition sub-zone, Eq. (\ref{eq:background_P}) is employed to regularize the particle distribution, whereas for other particles in the non-regularized transition sub-zone, no regularization technique is deployed. 

The activation status of the particles should be determined for the physical time evolution, and the operation of particle generation or particle deletion should also be implemented with regard to the activated blocks. The tag function $M$ denotes which zone the particle belongs to among the refinement zones, the transition zones and the non-refinement zones, for which they are marked $M=2$, $M=1$ and $M=0$ for the three zones, respectively. With the $M$ function, the activation status of the particles can be determined. By comparing to the $M$ function in the previous step, defined as $M^{\prime}$, the particle generation or particle deletion can be determined. The algorithm for determining the activation status  and particle generation or particle deletion is shown in Algorithm \ref{alg1}, where the symbol `$\gamma$' denotes the activation status, for which 1 and 0 represent `active' and `inactive', respectively.

\begin{algorithm}[t]
  \caption{Determination of the particle activation status and the corresponding particle generation or deletion}
  \label{alg1}
  \begin{algorithmic}[1]
  \FOR{all particles}

    \IF{$level=0$}
        \IF{$M=0$}      
            \STATE $\gamma \gets 1$;
        \ELSIF{$M=1$ and $M^{\prime}=0$}
            \STATE $\gamma \gets 1$;
            \STATE Generate refined particles by the splitting method;
        \ELSIF{$M=2$ and $M^{\prime}=0$}
            \STATE $\gamma \gets 0$;
            \STATE Generate refined particles by the splitting method; 
        \ELSIF{$M=1$ and $M^{\prime} \neq0$}
            \STATE $\gamma \gets 1$;
        \ELSIF{$M=2$ and $M^{\prime} \neq0$}
            \STATE $\gamma \gets 0$;
        \ENDIF

    \ELSIF{$level=1$}
        \IF{$M=0$}      
            \STATE Delete the corresponding particle;
        \ELSIF{$M=1$}
            \STATE $\gamma \gets 0$;
        \ELSIF{$M=2$}
            \STATE $\gamma \gets 1$;   
        \ENDIF
          
    \ENDIF
      
  \ENDFOR  
  \end{algorithmic}
\end{algorithm}
\subsubsection{Regularization of the newly generated particles in new blocks}
\label{sect:Regularization}

Once the new blocks are determined for refinement, new refined particles are generated. Usually, the newly generated particles do not have a smooth isotropic distribution, therefore, the distribution regularization of these particles is highly needed in order to ensure the simulation accuracy. Some particle repositioning methods, e.g., the particle diffusion methods \cite{lind2012incompressible}\cite{fu2019isotropic}\cite{ji2020consistent}\cite{fu2020adaptive} and the method based on the Voronoi particle dynamics \cite{fu2019optimal}, have been explored for this aim. In this work, the regularization method based on the particle diffusion will be employed.

In the previous sections, the activated blocks and the $M$ tag functions indicating zones are determined. For the $L_0$ particle with $M^{\prime}=0$ and $M=1$ or $2$, four new particles will be generated surrounding the coarse particle, similar to the standard APR method. In fact, the distribution of these newly generated particles is usually non-isotropic and non-smooth, especially between the newly refined particles and the old particles, which will inevitably affect the accuracy of the SPH simulations in the new block zones. Here, we propose a novel technique to eliminate the irregular particle distribution between the new blocks and other adjacent blocks. 

\begin{figure}[htb]
\centering
\includegraphics[width=0.7\textwidth]{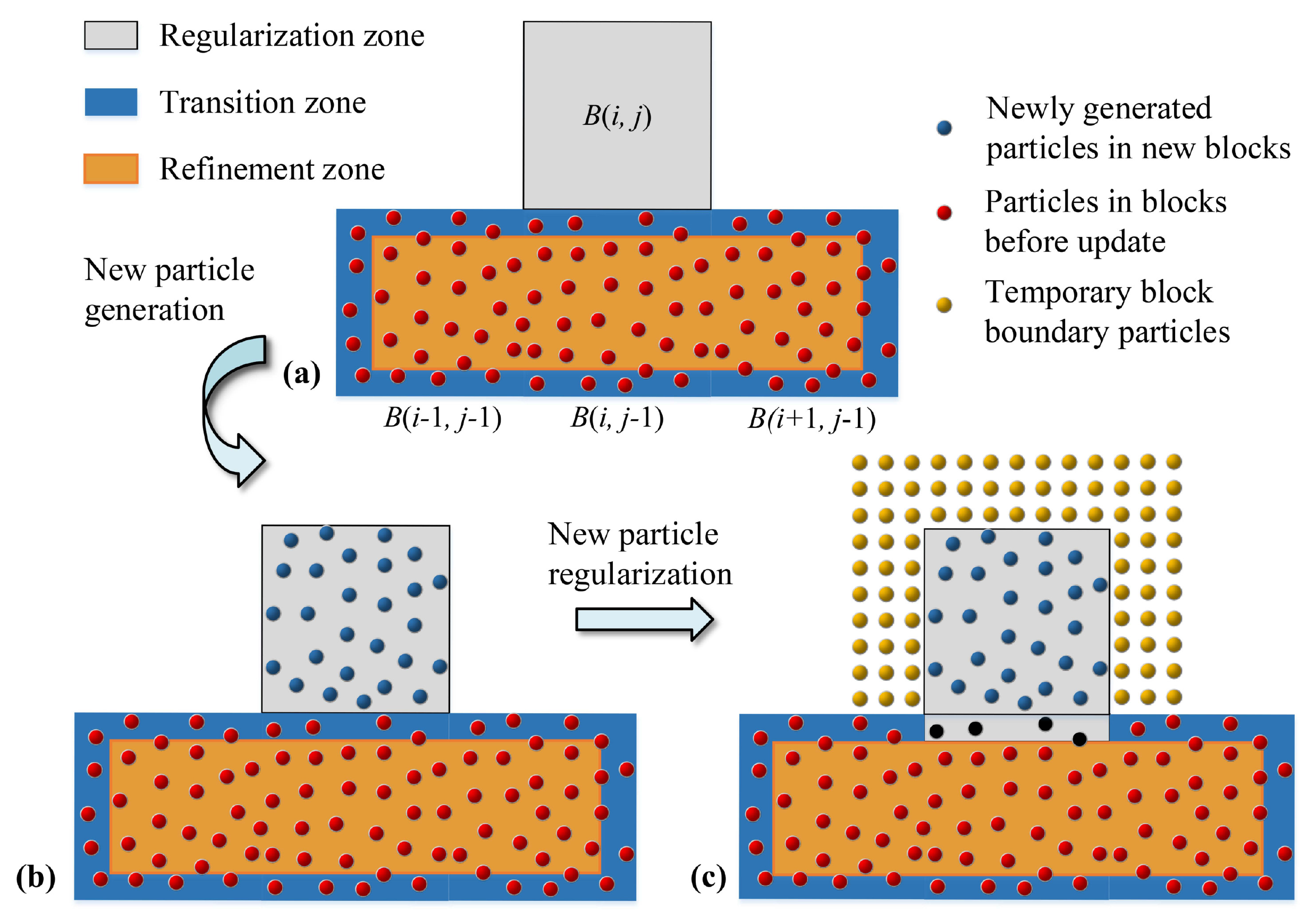}
\caption{Schematic of the particle regularization in the new blocks: for one new block $B(i,j)$ (gray), firstly the new particles are generated, then the particles in the new block and the particles in the linking area (from red to black) will be regularized. The temporary block boundary particles (yellow) will be deployed to constrain the regularized particles.}
\label{regularize}
\end{figure}

\begin{figure}[htb]
	\centering
	\includegraphics[width=0.7\textwidth]{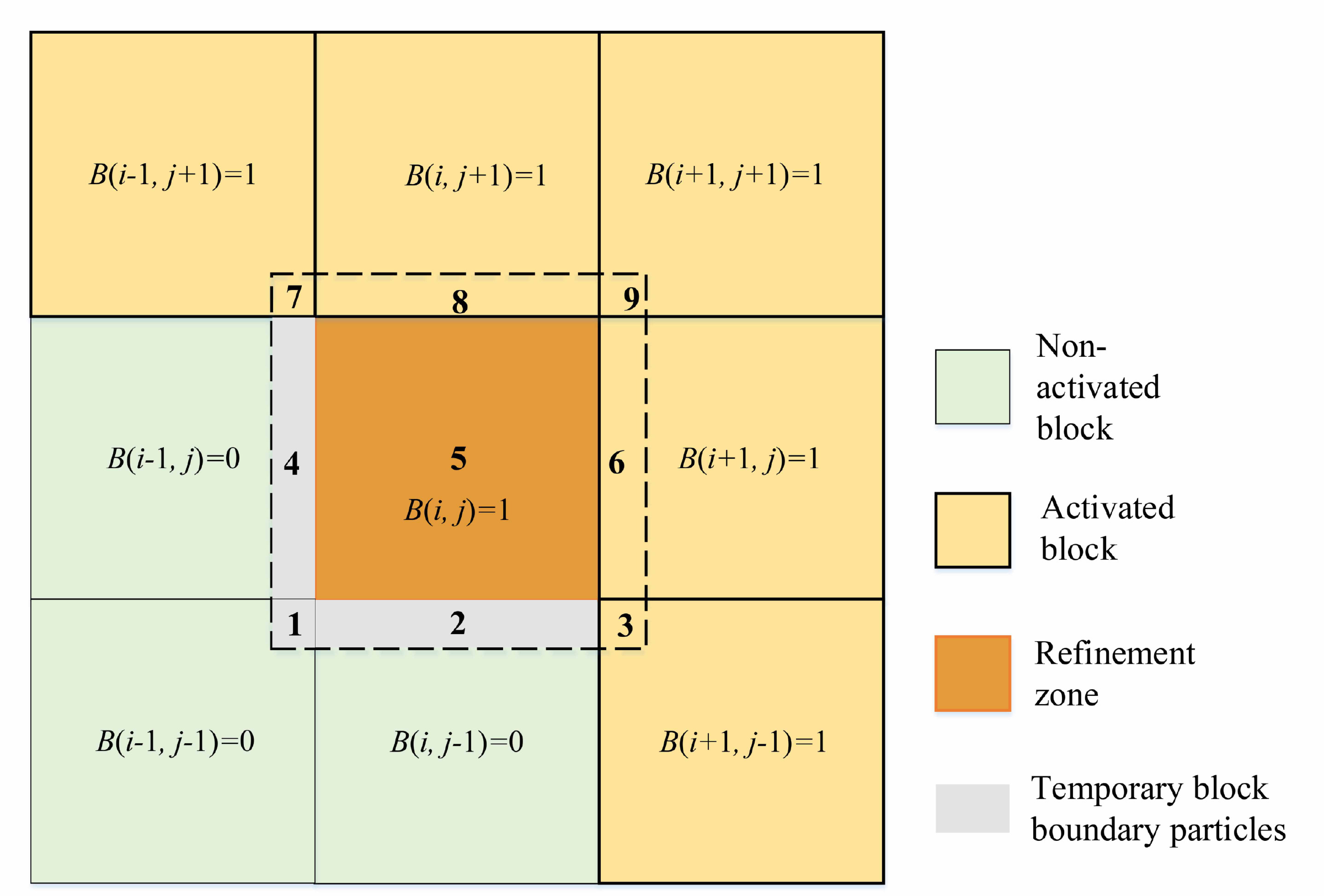}
	\caption{ {The generation of the temporary block boundary particles: for subdomains $1\sim 9$ around the targeted block $B(i,j)$, if the corresponding block that the subdomain locates in is non-activated, this subdomain will be filled with temporary block boundary particles, and vice versa.}}
	\label{temp_parti}
\end{figure}

Fig. \ref{regularize} illustrates the regularization technique for the new particles. For a block $B(i,j)=1$ with its previous status $B(i,j)=0$, it will be the targeted block for the new particle generation, as shown in Fig. \ref{regularize} (a). The $L_0$ particles in the gray block will generate new particles, marked in blue color, as shown in Fig. \ref{regularize} (b). Before the start of the SPH evolution, the positions of the new particles, in the gray area, will be regularized towards an isotropic distribution. In order to constrain the particle regularization inside the current block, the boundary particles are deployed around the block temporarily, as shown in Fig. \ref{regularize} (c).  {The generation technique of the temporary block boundary particles is shown in Fig. \ref{temp_parti}. The surrounding zone of block $B(i,j)$ is divided into nine subdomains $1\sim 9$, and if the corresponding block that the subdomain locates in is non-activated, this subdomain will be filled with temporary block boundary particles. Herein, the thickness of the temporary block boundary particles is $4\Delta x_1 + l_{tr}$, where $ l_{tr} $ is the thickness of the transition zone. } After the regularization, they are removed. It is worth noting that the black particles in the transition zone of the adjacent blocks will also be considered as the targeted particles for regularization, because the majority of these particles will change to `active' refined particles. As a result, the particles in the gray area will take part in the regularization in this figure. 

The regularization algorithm is based on Eq. (\ref{eq:background_P}). This equation provides the driving force for the particles with irregular distributions, and the particle positions $\boldsymbol{\bar {x}}$ in the regularization are then updated as
\begin{equation}
\label{eq:regulari}
\left\{
\begin{array}{l}
\begin{aligned}
&\Delta t = 0.25 \frac{h}{c},\\
&\boldsymbol{\bar{v}} =  \frac{\Delta t}{2 {\rho_0}} \boldsymbol{\nabla}  P^{\prime}_{ {B}i},\\
&\boldsymbol{\bar {x}}_{i+1} = \boldsymbol{\bar {x}}_{i} + \boldsymbol{\bar{v}} \Delta t.\\

\end{aligned}
\end{array}
\right.
\end{equation}
The regularization force $ \boldsymbol{\nabla}  P^{\prime}_{ {B}i} $ in Eq. (\ref{eq:regulari}) is determined according to whether the particle is in the surface area. The normal direction of surface particles are derived following \cite{antuono2010free}. For the particle in the surface zone, its normal direction $ \boldsymbol{n} $ is set equal to the normal direction of the nearest surface particle. Then,  {referring to the widely-used modification of background pressure gradient force for particles in the surface zone \cite{khayyer2017comparative}\cite{sun2017deltaplus},} the regularization force $ \boldsymbol{\nabla}  P^{\prime}_{ {B}i}$ can be obtained as
\begin{equation}
	\label{eq:regulari_force}
	\boldsymbol{\nabla}  P^{\prime}_{ {B}i}=\left\{\begin{array}{l}
		\boldsymbol{\nabla}  P_{ {B}i}- \boldsymbol{n} (\boldsymbol{n} \cdot \boldsymbol{\nabla} P_{ {B}i}), \text {  for particles in the surface zone}, \\
		\boldsymbol{\nabla}  P_{ {B}i}, \text {  otherwise}.
	\end{array}
	\right.
\end{equation}
Eq. (\ref{eq:regulari}) is iterated until the maximum norm of the driving force in the computational domain is lower than a threshold, as
\begin{equation}
	\label{eq:thresd}
	\dfrac{\left| \nabla  P^{\prime}_{ {B}i} \right|_{max} h}{\rho_{0} c^{2}} \le \epsilon_{th} ,
\end{equation}
where $\epsilon_{th}=5 \times 10^{-6} $ is adopted with the Gaussian kernel deployed in this paper. 
When this criterion is satisfied, the desired isotropic distribution is considered to be achieved. When the new particles are regularized to the new positions in the new block, the initial field values should be assigned. The Shepard method is implemented by interpolating from the surrounding `active' particles, as suggested in the standard APR method \cite{chiron2018analysis}. After the field value assignment, all the particles in the computational domain will take part in the physical time evolution by solving the SPH governing equations.  {It is worth noting that section \ref{sect:Regularization} is only used to regularize the newly generated particles when new blocks are activated. }

To summarize the BAPR method, an overview of  the BAPR method is shown in Algorithm \ref{alg2}. Firstly, the activation status of the block array is determined from line 1 to line 8. From line 9 to line 21, the tag function $M$ distinguishing the transition zone or the refinement zone for each particle is obtained. In line 22, the activation status of each block is obtained based on the $M$ value. From line 23 to line 35, the position regularization for new particles is implemented to achieve an isotropic particle distribution. After that, the initial field values are assigned with a Shepard interpolation method in line 36. 

Moreover, in this work, the multi-resolution framework is based on the parallel MPI implementation. An efficient multi-resolution data structure for the neighbor searching and the parallel communication is employed in this framework following \cite{fu2019parallel}\cite{fu2017physics}\cite{ji2019new}\cite{ji2019lagrangian}. More technical details are given in \cite{ji2019new}.

\begin{algorithm}
  \caption{Overview of the BAPR method }
  \label{alg2}
  \begin{algorithmic}[1]

  \FOR{all particles}
  \IF{the tracked quantity of interest $\phi$ reaches the threshold ${\phi}_{th}$  \OR
    the phase is solid}
  \STATE Determine the targeted block index ($i,j$) in the block array;
  \STATE Determine the corresponding subdomain $S$ in the block $B(i,j)$;
  \STATE Determine the neighbor block $B(i^{\prime},j^{\prime})$ of subdomain $S$ referring to Fig. \ref{BAPR}(b);  
  \STATE $B(i^{\prime},j^{\prime}) \gets 1$;
  \ENDIF
  \ENDFOR  
  
  \FOR{all particles}
    \STATE Determine the targeted block index ($i,j$) in the block array;
      \STATE Determine the corresponding subdomain $S$ in the block $B(i,j)$;
      \IF{$B(i,j)=1$}
          \IF{all adjacent blocks of the subdomain $S$ are activated}
            \STATE $M \gets 2$;
          \ELSE
            \STATE $M \gets 1$;
          \ENDIF
      \ELSIF{$B(i,j)=0$}
        \STATE $M \gets 0$;
      \ENDIF
      
  \ENDFOR    

  \STATE Determine the particle activation status with Algorithm \ref{alg1};
  
  \FOR{all $Block(i,j)$}
  \IF{$Block(i,j)=1$ and  $Block(i,j) (\text{before update})=0$}
  \STATE Identify the zone for temporary block boundary particles;
  \STATE Deploy the temporary block boundary particles;
  \STATE Map the temporary block boundary particles to the local data structure;
  \STATE Determine all the particles for regularization according to their positions;
  \WHILE{the regularization criterion is not satisfied}
    \FOR{all the particles for regularization}
        \STATE Implement the iteration process based on Eq. (\ref{eq:regulari});  
    \ENDFOR
  \ENDWHILE

  \ENDIF
  \ENDFOR
  \STATE Assign field values for the regularized particles with the Shepard interpolation method;
  \end{algorithmic}
\end{algorithm}

\section{Numerical validations}
\label{section:4}

In this section, several cases are presented using the proposed multi-resolution SPH framework. The cases include water impact on an elastic beam, dam-breaking through an elastic beam, flapping beam induced by the flow over a cylinder, flow over an inclined elliptical cylinder, and body entry problems.

\subsection{Water impact on an elastic beam}

\begin{figure}
\centering
\includegraphics[width=0.6\textwidth]{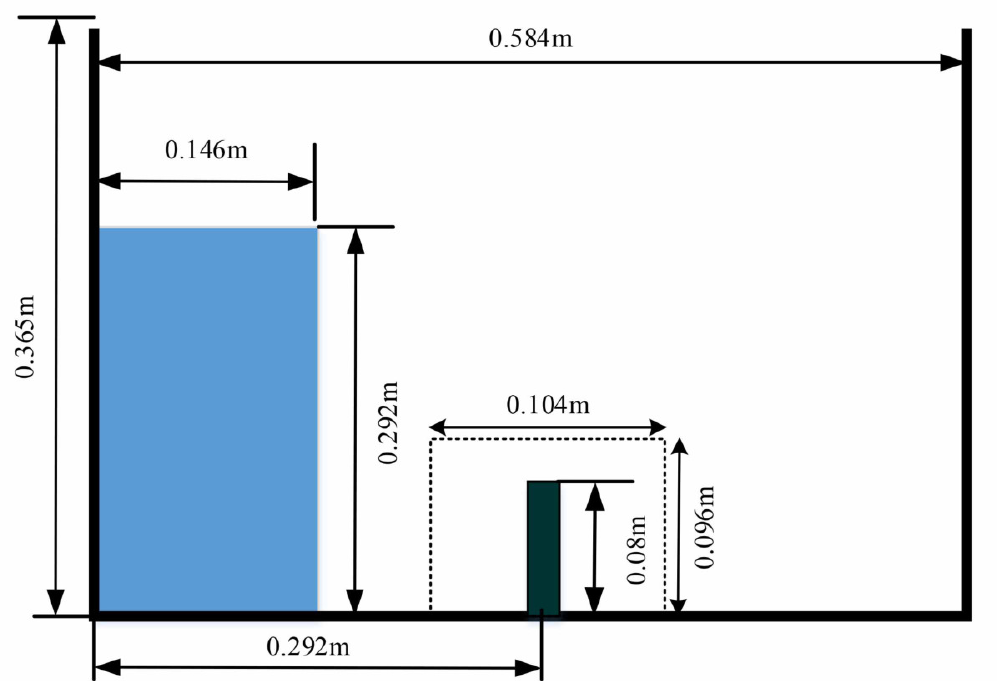}
\caption{Schematic of the water impact on the elastic beam. }
\label{Idelsohn}
\end{figure}

Fig. \ref{Idelsohn} shows the schematic of the FSI problem of water impact on an elastic beam, where the water breaks from the dam and then flows onto the elastic beam. This problem has been investigated by Idelsohn et al. \cite{idelsohn2008unified} with the Particle Finite Element Method (PFEM). The water domain is 0.292 m in height and 0.146 m in width. The length of the elastic beam is 0.08 m and the thickness $T$ is 0.012 m, with density $\rho_S=2500$ $\rm kg/m^{3}$, Young's modulus $E=10^{6}$ $\rm Pa$ and Poisson's ratio $\nu=0$. The water is considered as inviscid fluid with the density as $\rho=1000$ $\rm kg/m^{3}$. For this case, the present APR method is firstly deployed.  {$ \Delta x_{1}=T/8 $ is deployed for the refinement resolution}, and the APR zone in this case is deployed within the dashed-line box shown in Fig. \ref{Idelsohn}, where the beam is located in the center.  {Considering that this is a typical FSI case with violent flows, the shifting velocity constraint Eq. (\ref{eq:dv_con}) is used}.  {In this case, an hourglass control algorithm proposed by Ganzenmüller et al. \cite{ganzenmuller2015hourglass} is used to eliminate the tensile instability in the beam.}

\begin{figure}
\centering
\includegraphics[width=0.8\textwidth]{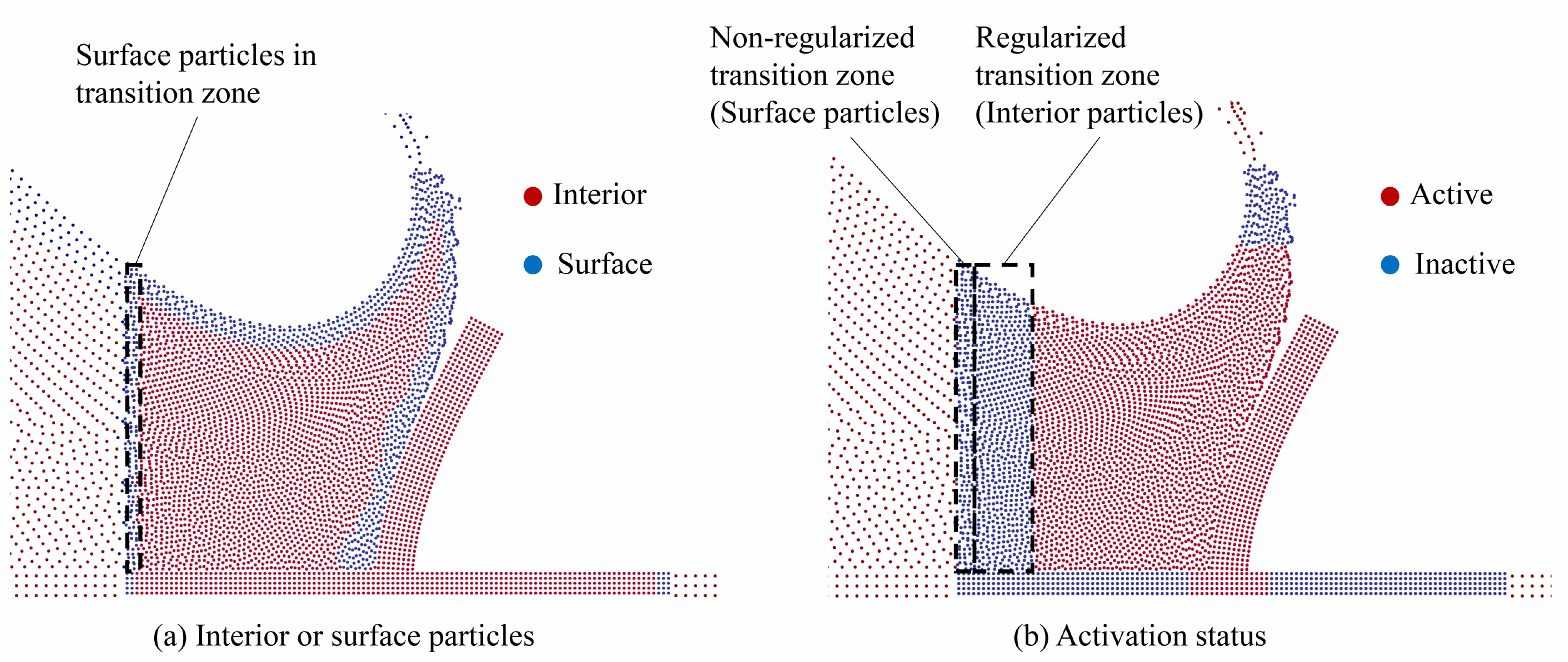}
\caption{ {Illustration of the interior or the surface particles,  and the activation status in the improved APR method.} }
\label{inter_acti}
\end{figure}

In order to further illustrate the regularization technique in the present modified APR method, a simulation snapshot is shown in Fig. \ref{inter_acti}. In Fig. \ref{inter_acti} (a), blue particles are surface particles and red are interior particles. The regularization technique is not deployed for surface particles. If the surface particles are in the transition zone, these particles will be in the non-regularized transition zone at the outer layer, as shown in Fig. \ref{inter_acti} (b). The blue refined particles are the `inactive' transition particles, which are used as buffer particles of the red `active' particles. For those in the transition zones, at the outer layer, the particles will be identified as surface particles and the regularization strategy will not be implemented; for those in the inner layer, the particles are identified as interior particles and particle regularization will be implemented. This method ensures that the buffer particles in the transition zones have a smooth transition to the `active' particles.

\begin{figure}
\centering
\includegraphics[width=0.8\textwidth]{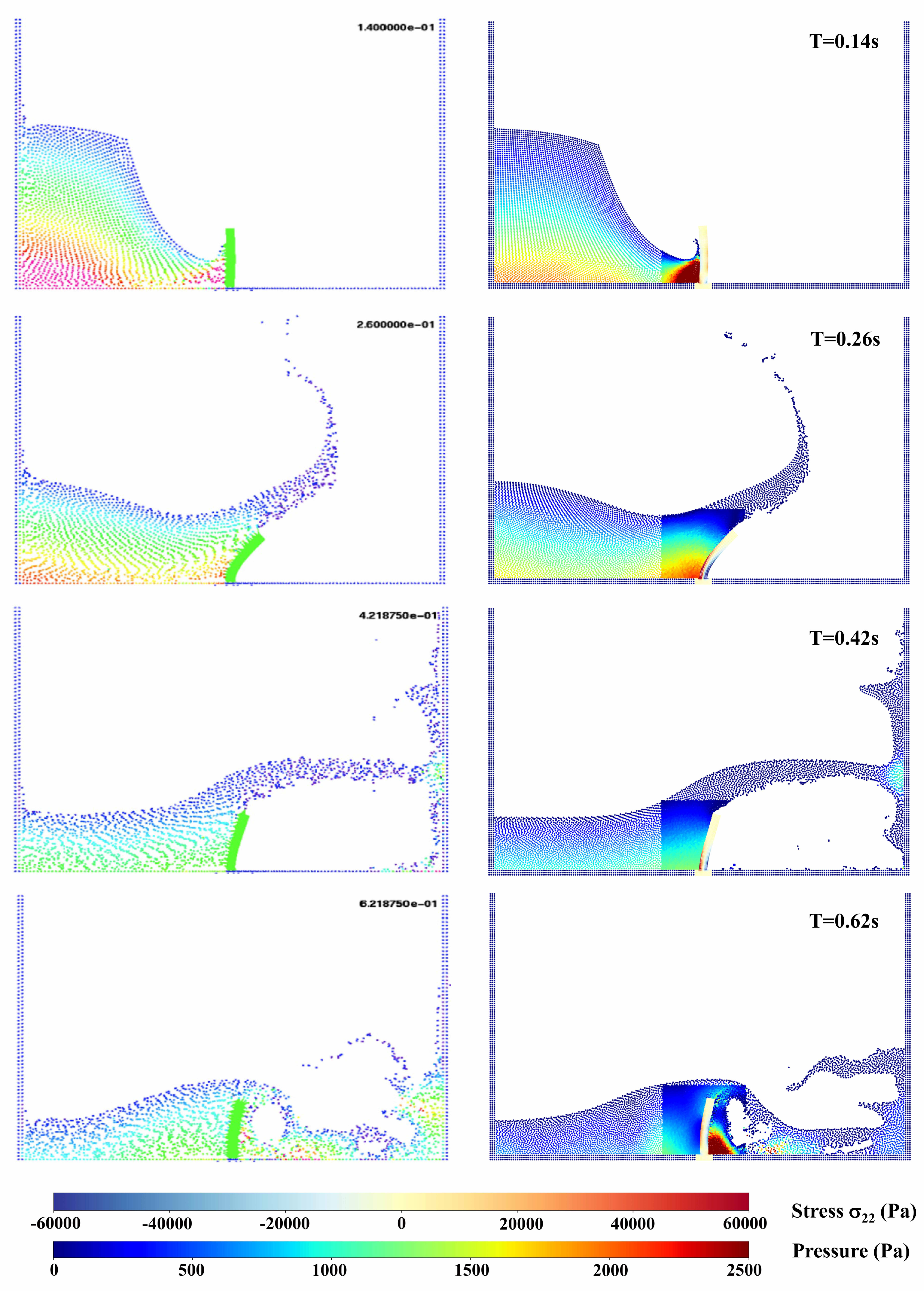}
\caption{ {Water impact on an elastic beam: simulation snapshots with the present multi-resolution SPH method (right) and the comparisons with those obtained by the PFEM method \cite{idelsohn2008unified} (left).} }
\label{dam_imp_reslts}
\end{figure}

\begin{figure}
\centering
\includegraphics[width=0.7\textwidth]{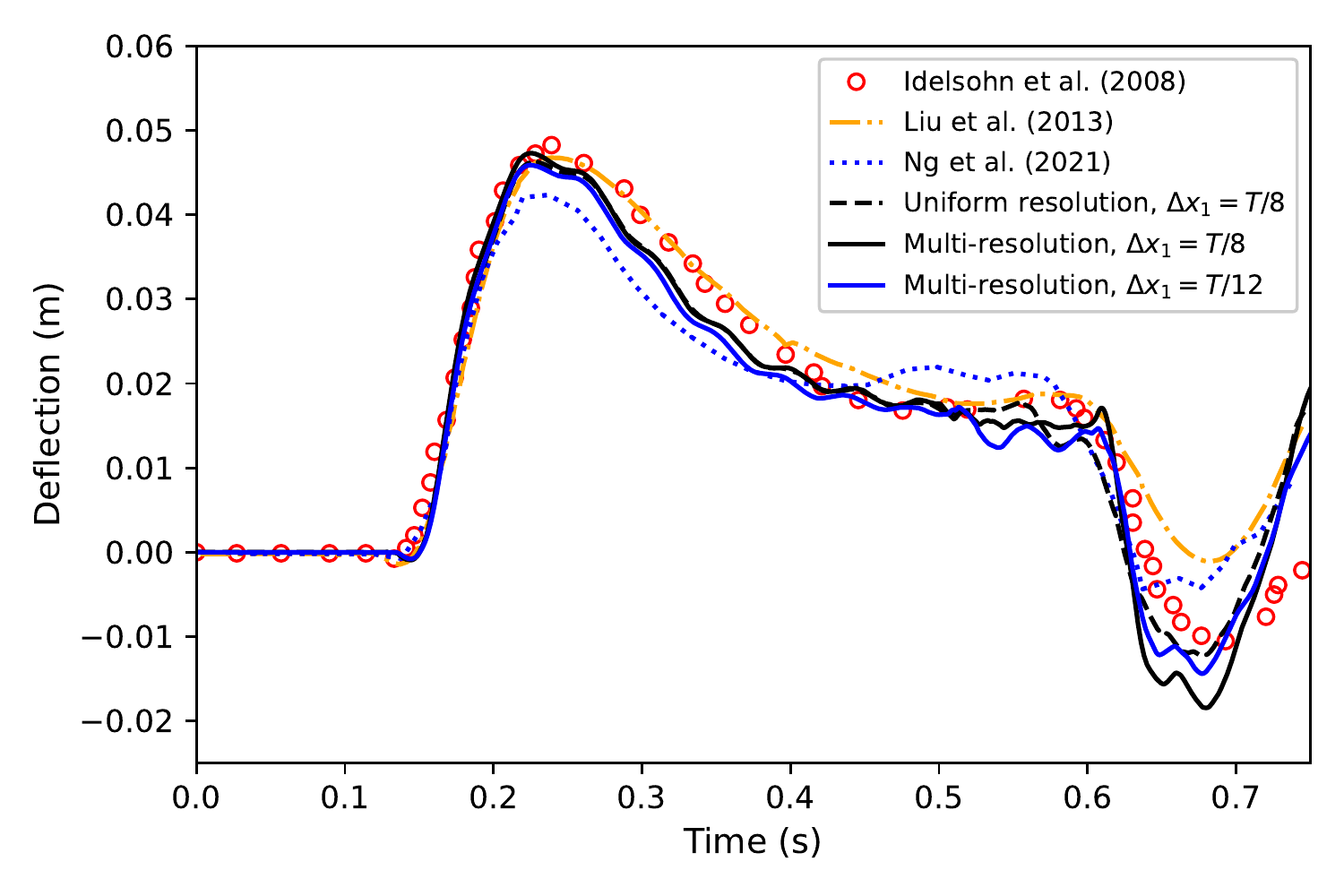}
\caption{Deflection statistics of the elastic beam tip with the modified APR method, and the comparisons with those from the uniform-resolution simulation and the simulation results by Idelsohn et al. \cite{idelsohn2008unified}, Liu et al. \cite{liu2013numerical} and Ng et al. \cite{ng2022improved}  {(resolution $ \Delta x_1=T/12 $)}.}
\label{dam_plot}
\end{figure}

The simulation snapshots of this case with the APR method is shown in Fig. \ref{dam_imp_reslts}, where the results are also compared with Idelsohn's simulation \cite{idelsohn2008unified} using PFEM. It is observed that the pressure profile of the present simulation results using our multi-resolution method has a great agreement with that using PFEM. Owing to the ALE nature of the present SPH method and the particle regularization technique in the transition zones, the particle distribution of the present method is more regular than using PFEM, and the pressure field between the refinement zones and the coarse-resolution zones is satisfactorily smooth. Fig. \ref{dam_plot} plots the deflection of the beam and its comparison with the results by Idelsohn et al. \cite{idelsohn2008unified}, Liu et al. \cite{liu2013numerical} and Ng et al. \cite{ng2022improved}. In this figure, it is shown that the present result almost overlaps with that using the uniformly fine resolution before $t=0.4$ s; after that, there is a little deviation, which can be also observed in other results shown in the figure. In fact, this deviation is under expectation for this case because the flow impact is violent, and the result is very sensitive to the particle distribution \cite{ng2022improved}. Though there is a slight oscillation, the results from the present multi-resolution method still agree quite well with other studies.  {Moreover, a resolution of $\Delta x=T/12$ (same as that in Ng et al. \cite{ng2022improved}) is deployed to further validate this case, and it is shown that the present simulations with different resolutions are consistent with each other and agree well with other studies.}
Therefore, we conclude that the present multi-resolution method is effective to improve the local accuracy with local particle refinement.

\subsection{Dam-breaking flow through an elastic beam}

\begin{figure}
\centering
\includegraphics[width=0.6\textwidth]{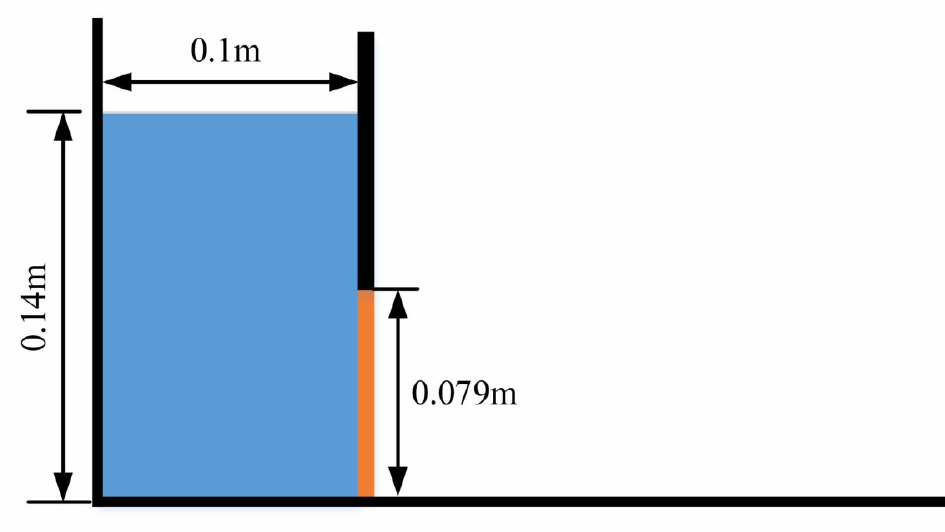}
\caption{ {Schematic of the dam-breaking flow through an elastic beam.}}
\label{dam_schm}
\end{figure}

In this case, the dam-breaking flow though an elastic beam is simulated with the BAPR method. This case was firstly studied by Antoci et al. \cite{antoci2007numerical} by experiment. Initially, the water is constrained by the tank and the elastic beam, with its upper-half clamped and lower-half free. Upon the start of simulation, the water will flow out from the lower end of the elastic beam. The elastic beam is made of rubber, and the linear elastic constitutive relationship is employed for the material, with density $\rho=1100$ $\rm kg/m^{3}$, Young's modulus $E=7.8$ $\rm MPa$ and Poisson's ratio $\nu=0.47$, following the setup of Zhang et al. \cite{zhang2021multi}. The water is considered as inviscid fluid with the density as $\rho=1000$ $\rm kg/m^{3}$. The setup of this case is shown in Fig. \ref{dam_schm}, where the water column is $0.1$ $\rm m$ in the width and 0.14 $\rm m$ in the height, and the thickness of the beam is $T=0.005$ $\rm m$.  In this case, the proposed BAPR method is employed, and the targeted characteristic for BAPR method is the solid phase,  {the activated blocks around the beam are identified adaptively with the method described in Fig. \ref{BAPR}(b)}.  {The refined resolution is $\Delta x_{1}=T/8$. }   {The block sizes in this case are $L_x=0.025 \ \rm m$ and $L_y=0.025 \text{ } \rm m$ in the horizontal and vertical directions, respectively. }

\begin{figure}
\centering
\includegraphics[width=0.8\textwidth]{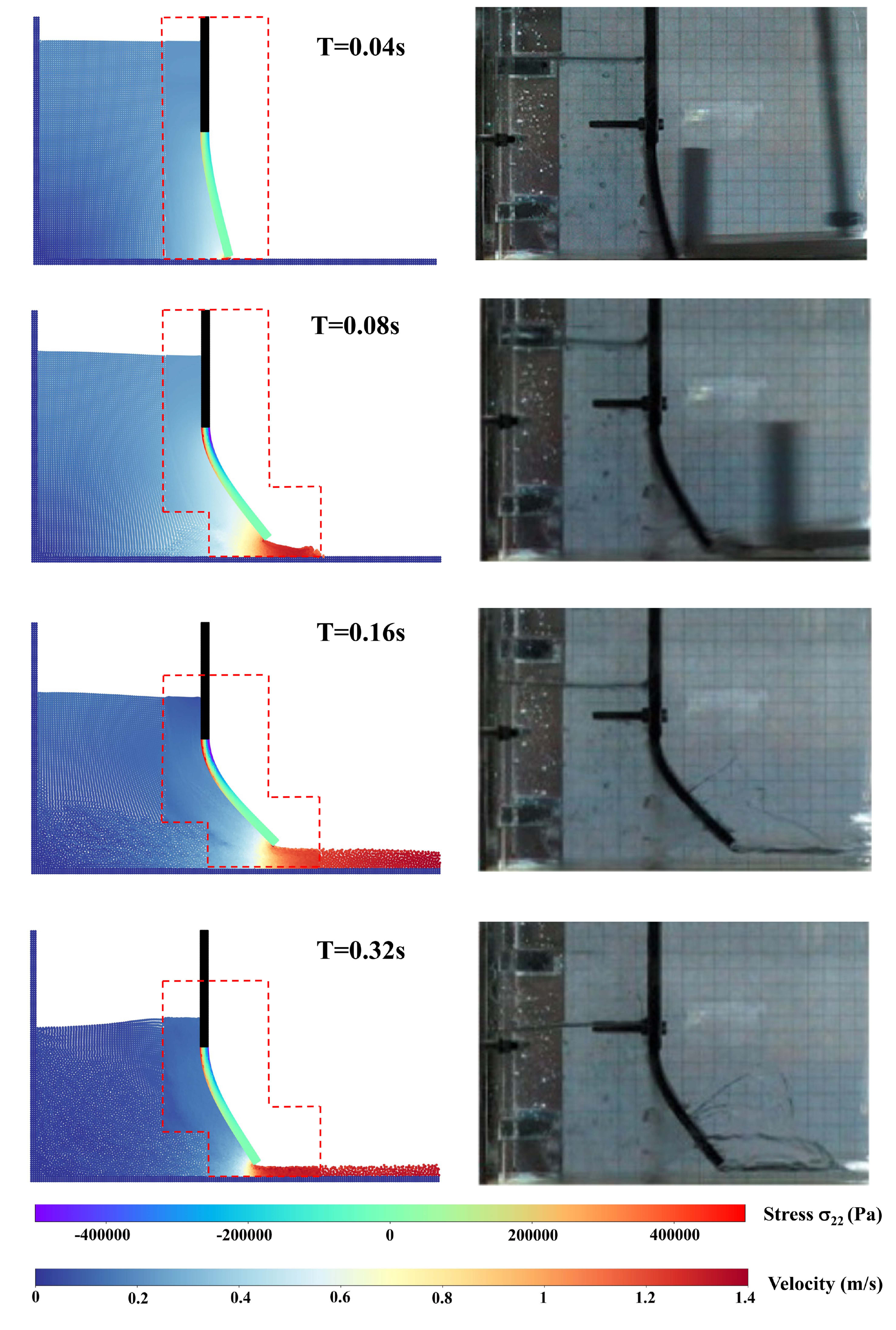}
\caption{Dam-breaking flow through an elastic beam: distributions of the fluid velocity and the solid stress component $\sigma_{22}$ at different physical time (left), and the comparisons with the experiments by Antoci et al. \cite{antoci2007numerical} (right).}
\label{dam_snap}
\end{figure}

\begin{figure}
\centering

\subfigure{
\begin{minipage}{1\textwidth}
\centering
\includegraphics[width=0.7\textwidth]{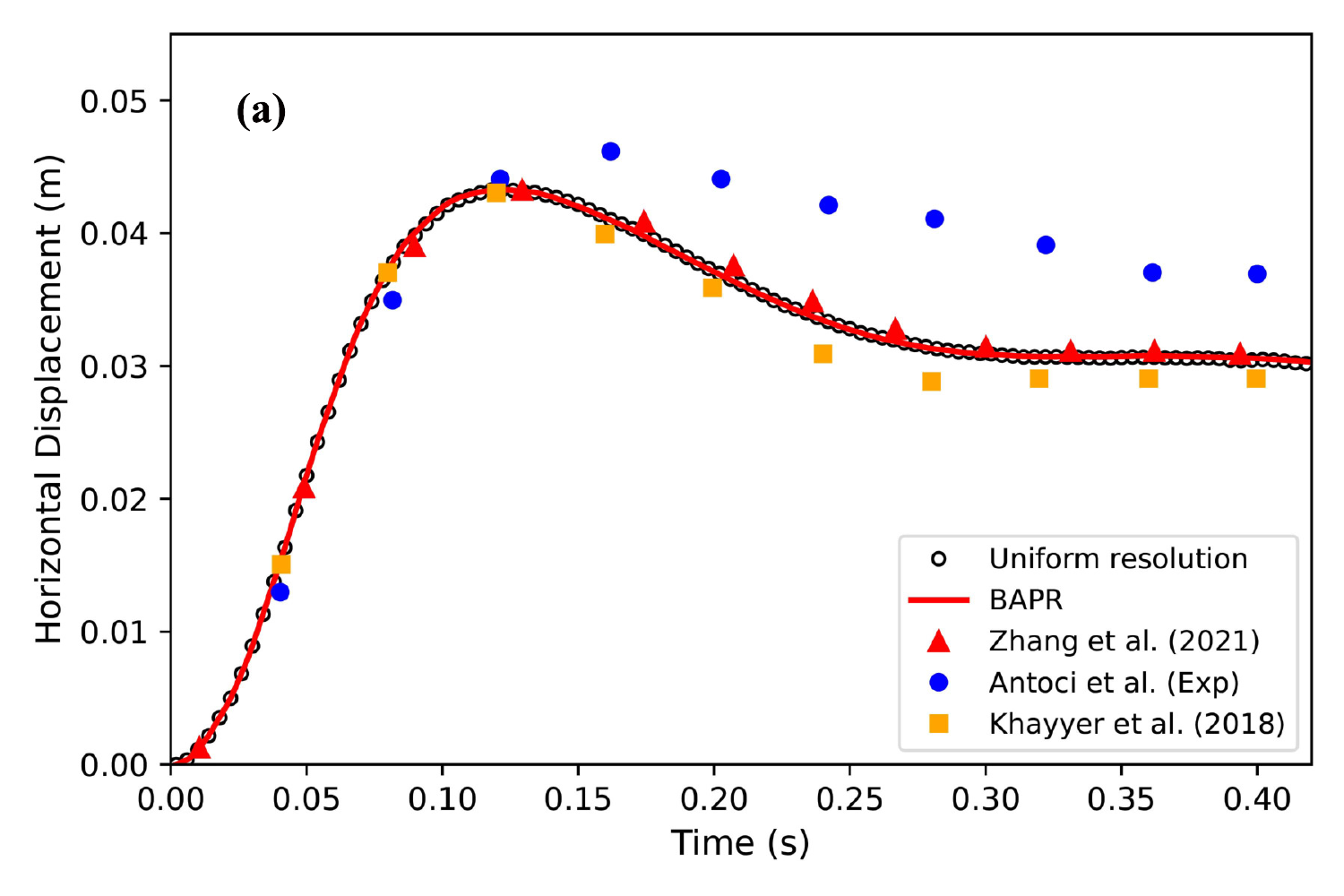}
\end{minipage}
}
\subfigure{
\begin{minipage}{1\textwidth}
\centering
\includegraphics[width=0.7\textwidth]{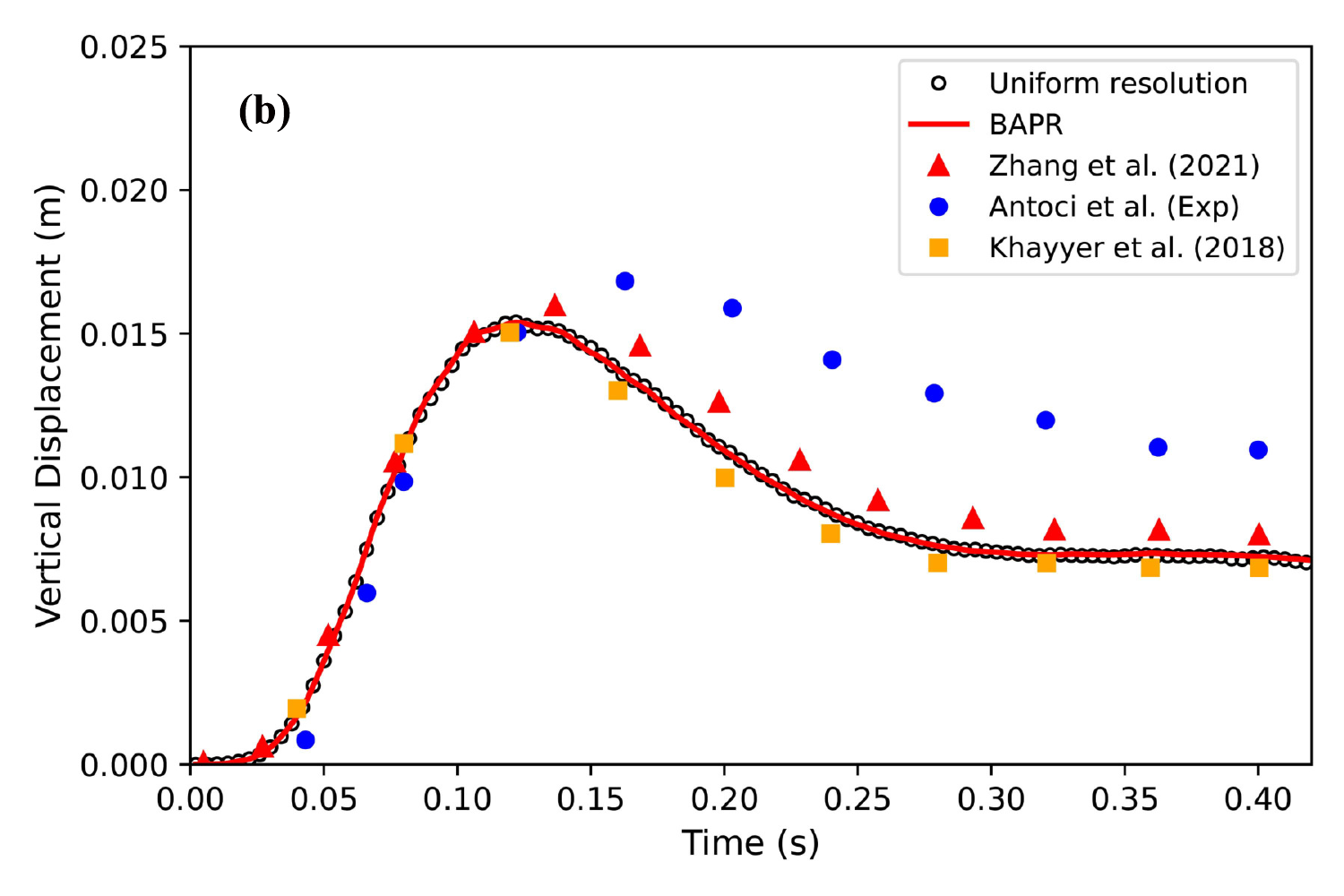}
\end{minipage}
}

\caption{Statistics of the horizontal (a) and vertical (b) displacements of the beam tip, and the comparisons with the  experimental data by Antoci et al. \cite{antoci2007numerical}, and the simulation results by Zhang et al. \cite{zhang2021multi} and Khayyer et al. \cite{khayyer2018enhanced}   {($E=12 \ \rm {MPa},  \nu=0.4$)}.}
\label{dam_plt}
\end{figure}

Fig. \ref{dam_snap} shows the snapshots of simulation results with the BAPR method, in which the velocity field is shown, and the area enclosed with the red dashed lines denotes the refinement BAPR zone. At time $t=0.04$ $\rm s$, the deformation of the beam is small, and the refinement blocks do not change in topology. At time $t=0.08$ $\rm s$, the beam deforms intensely, and the refinement blocks track the beam adaptively. From time $t=0.08 \sim 0.32$ $\rm s$, the beam still stays in the refinement blocks, therefore these block zones do not change in topology. Throughout the process, it is shown that our results have a very good agreement with the experimental data by Antoci et al. \cite{antoci2007numerical}. Fig. \ref{dam_plt} illustrates the quantitative comparison between different results. Firstly, it is shown that the results by the new BAPR method and the uniform resolution method almost entirely overlap with each other, which demonstrates that the BAPR method is capable of achieving a higher resolution with an adaptive local refinement. The present results also agree well with those by Zhang et al. \cite{zhang2021multi} and Khayyer et al. \cite{khayyer2018enhanced}  {($E=12 \ \rm {MPa},  \nu=0.4$)}. The little discrepancy between the present simulation results and the experiments may come from the material mechanical descriptions, e.g., the deployment of the constitutive relationship and the neglect of the nonlinear behaviors. 

\subsection{Flapping beam induced by the flow over a cylinder}
\label{sect:Flapbeam}

\begin{figure}
\centering
\includegraphics[width=0.7\textwidth]{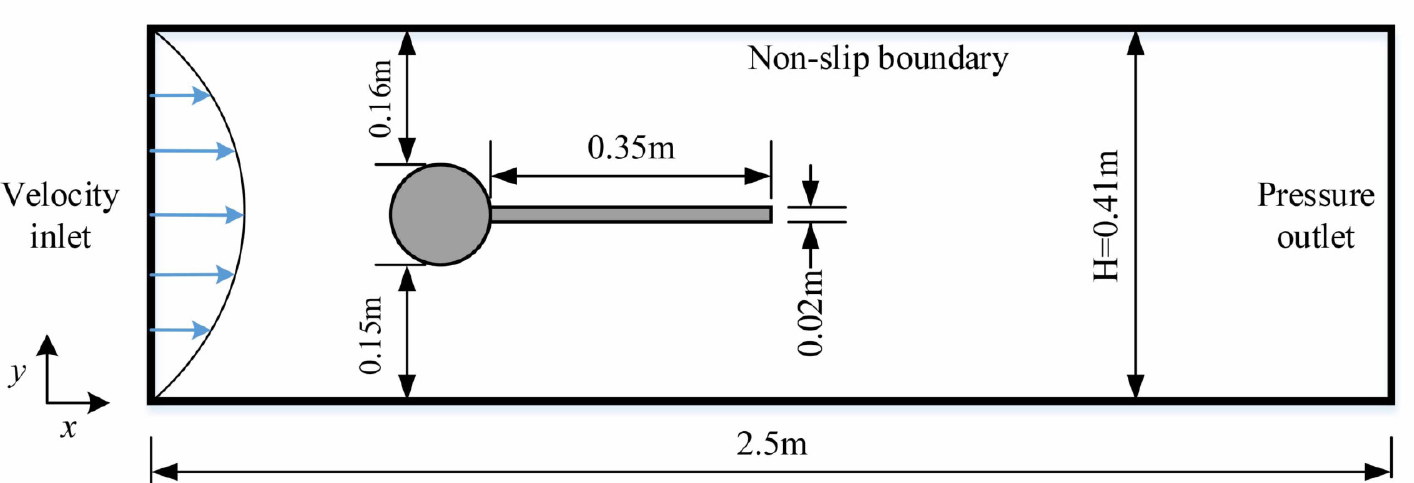}
\caption{Schematic of the flapping beam induced by the flow over a cylinder. }
\label{flapping}
\end{figure}

The flow induced vibration of a beam behind a cylinder is simulated with the proposed BAPR method. In this case, the flexible beam is attached to the end of a cylinder, and the beam will vibrate with a self-sustaining frequency after the onset of the instability. The benchmark setup following \cite{turek2006proposal} is shown in Fig. \ref{flapping}, where the material parameters of the beam are given as: density $\rho_S=10000$ $\rm kg/m^{3}$, Young's modulus $E=1.4\times10^{6}$ $\rm Pa$ and Poisson's ratio $\nu=0.4$, respectively.  {The density of the fluid is $\rho_F=1000$ $\rm kg/m^{3}$}, and its kinetic viscosity is $0.001$ $\rm m^{2}/s$. The mean inlet velocity is $U_{0}=1$ $\rm m/s$, and the velocity profile obeys
\begin{equation}
\label{eq:inlet}
\begin{array}{l}
\begin{aligned}
v(y)=1.5\bar{U}_{t}(H-y)y/H^{2},
\end{aligned}
\end{array}
\end{equation}
where $\bar{U}_{t}$ is introduced to ensure a smooth transition from 0 to the steady velocity $U_{0} $, and expressed as
\begin{equation}
\label{eq:inlet1}
\bar{U}_{t}=\left\{\begin{array}{l}
0.5U_{0}(1-cos(0.5\pi t)),\  t\textless 2, \\
U_{0}, \ t\ge 2.
\end{array}
\right.
\end{equation}

The pressure of the inlet is derived with the Shepard interpolation from the nearest SPH particles. For the outlet, the pressure is set as zero according to \cite{long2021coupling}. Likewise, the velocity is derived with the Shepard interpolation from the nearest SPH particles. A buffer zone with a thickness of $6\Delta x_{0}$ is allocated at the inlet and the outlet, where $\Delta x_{0}$ denotes the particle spacing of the coarse level. For the inlet particles, if particles move with a displacement of $\Delta x_{0}$ inwards away from the inlet line, new particles are created; for the outlet particles, if particles move with a displacement of $\Delta x_{0}$ outwards away from the computational domain boundaries, then they are deleted from the computational domain.  {The refined resolution is $ \Delta x_{1}=T/8$, where $ T $ is the thickness of the beam.} In this case, the BAPR adaptation criterion is based on the velocity characteristic, with which the high velocity area is refined accordingly.  {Here, the velocity threshold is 
defined as $ \left| \boldsymbol{v} \right|_{th} = 1.7 \text{ }\rm {m/s} $.}  {The block array sizes $(n_x,n_y)$ in this case are $n_x=25$ and $n_y=5$ in the horizontal and vertical directions, respectively. }

\begin{figure}
	\centering
	\includegraphics[width=1\textwidth]{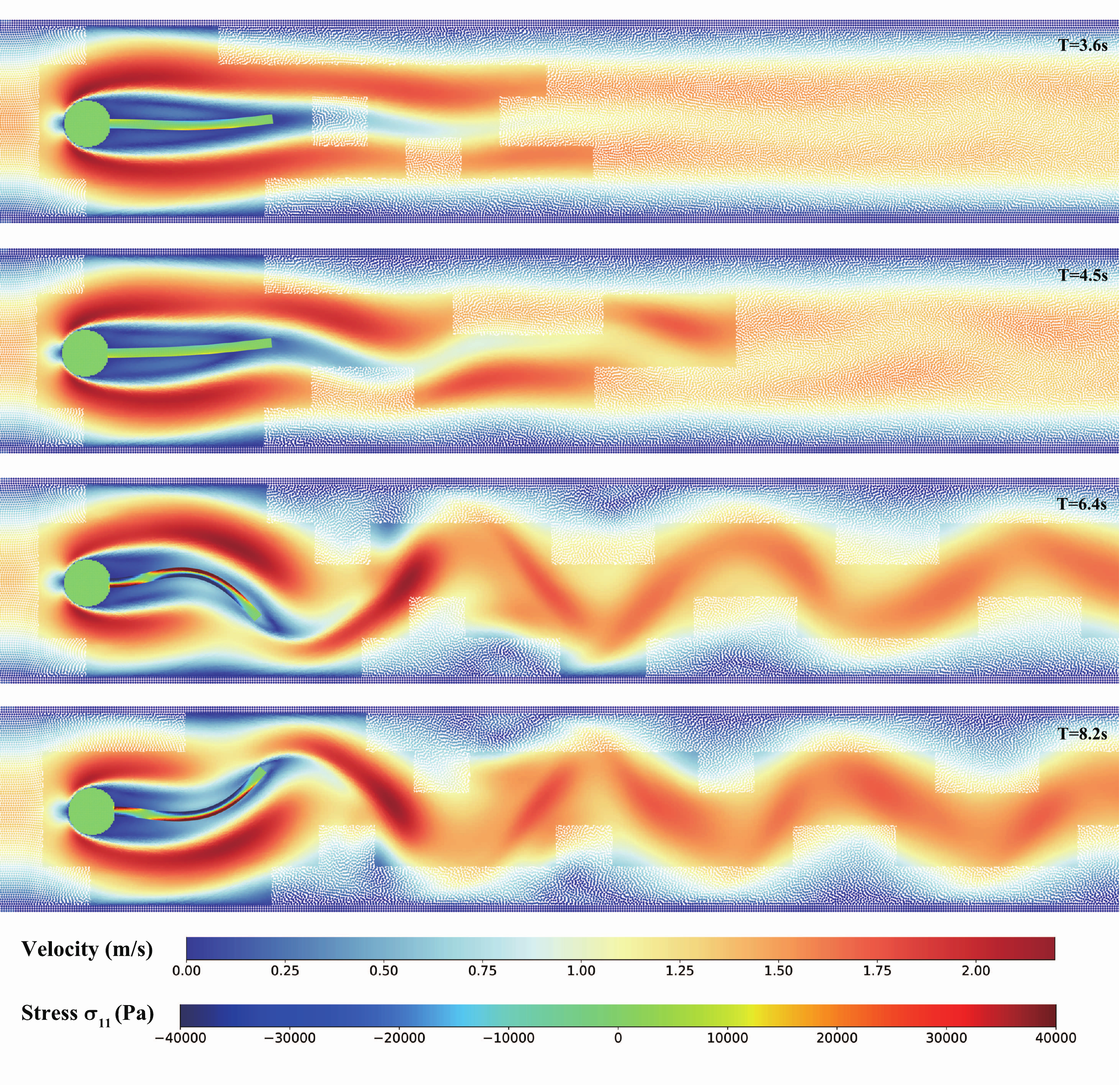}
	\caption{Snapshots of the velocity distributions of the flapping beam induced by the flow over a cylinder at different physical time. The refinement blocks (dark-color regions) are identified based on the velocity characteristic with the proposed BAPR method.}
	\label{flapbeam_bapr}
\end{figure}

\begin{figure}[htb]
\centering
\includegraphics[width=0.7\textwidth]{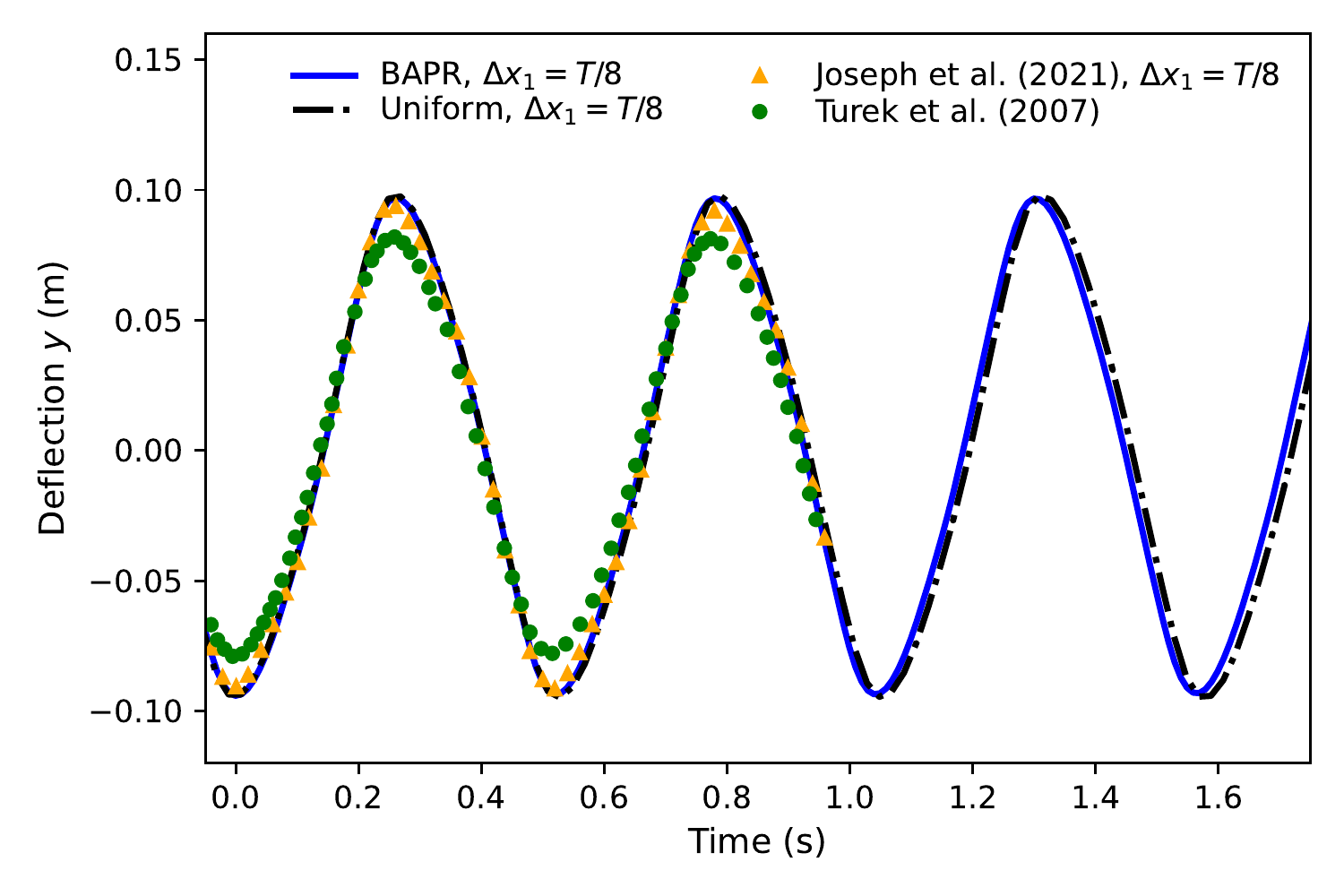}
\caption{ {Deflection statistics of the elastic beam tip and the comparison with the results from Joseph et al. \cite{o2021fluid} and Turek and Hron \cite{turek2006proposal}.} }
\label{flapping_plt}
\end{figure}

In Fig. \ref{flapbeam_bapr}, the simulation results of the flapping beam at different time are given. It is shown that the high velocity region is tracked and refined adaptively throughout the process, and the simulation field data in the refinement zones has a smooth transition to that in the coarse resolution regions due to the particle regularization technique used in the transition zones and newly generated particles. The deflection of the beam tip with time is shown in the Fig. \ref{flapping_plt}, where results from other studies are compared, and all results are shifted in phase for fair comparisons. It can be seen that the results from the BAPR method and the uniform resolution simulation almost overlap with each other, which demonstrates the accuracy of the proposed BAPR method. Also, the results from the BAPR method agree well with the SPH results by Joseph et al. \cite{o2021fluid} using the same resolution $ \Delta x_{1}=T/8$. The marginal difference of the amplitude with results from \cite{turek2006proposal} using the implicit FEM solver with an ALE formulation is due to the different choices of the damping effects, and no damping is employed for the solid phase in this case. The present resolved vibration amplitude is quite close to the result $0.092$ $\rm{m}$ obtained by Bhardwaj and Mittal \cite{bhardwaj2012benchmarking} where no damping is employed. Table \ref{tb:deflection} shows the comparison between different simulation results. It is shown that the vibration amplitude and period of the present simulation have a reasonable agreement with results of other studies.  {Table \ref{tb:flpbeam_t} shows the comparison of the average computational time for simulating each 0.02\ $\rm s$ between the uniform resolution simulation and the proposed BAPR method. In this comparison, one MPI rank with 32 threads is employed for the simulations, and a large reduction of computational costs is observed with the present BAPR method.}
Overall, the BAPR method is capable of tracking complex flow structures by refining the targeted zones adaptively, while achieving the same accuracy as the uniform resolution simulation.

\begin{table}
\renewcommand{\arraystretch}{1.3}
\caption{{Comparison between statistics of the flapping beam induced by the flow over a cylinder} }
\begin{center}
\begin{tabular}{p{6cm} p{4cm} p{4cm} }
\toprule [1.2 pt]
Cases & Amplitude (m) & Period (s)\\
\hline
Turek and Hron \cite{turek2006proposal} & 0.083 & 0.526  \\
Bhardwaj and Mittal \cite{bhardwaj2012benchmarking} & 0.092 & 0.526  \\

Joseph et al. \cite{o2021fluid}, $ {\Delta x_1=T/16}$ & 0.078 & 0.5299  \\
Joseph et al. \cite{o2021fluid}, $ {\Delta x_1=T/8}$ & 0.093 & 0.5234  \\
Zhang et al. \cite{zhang2021multi}, $ {\Delta x_1=T/8}$ & 0.086 & 0.5291  \\
Sun et al. \cite{zhang2021multi}, $ {\Delta x_1=T/10}$ & 0.088 & 0.5341  \\
BAPR, $ {\Delta x_1=T/8}$ & 0.094 & 0.5230  \\
\bottomrule [1.2 pt]
\end{tabular}
\end{center}
\label{tb:deflection}
\end{table}

\begin{table}
	
	\renewcommand{\arraystretch}{1.2}
	\caption{ {Comparison of the average computational time for simulating each 0.02\ $\rm s$ between the uniform resolution simulation and the proposed BAPR method.} }
	\begin{center}
		\begin{tabular}{p{6.5cm} p{2cm} p{4cm} }
			\toprule [1.2 pt]
			Cases &   BAPR & Uniform resolution\\
			\hline
			$ {\Delta x_1=T/8}$, 1 MPI (32 threads) & 62 $\rm s$ & 110 $\rm s$  \\
			$ {\Delta x_1=T/12}$, 1 MPI (32 threads) & 191 $\rm s$ & 451 $\rm s$  \\				
			\bottomrule [1.2 pt]
		\end{tabular}
	\end{center}
	\label{tb:flpbeam_t}
\end{table}

\subsection{Flow over an inclined elliptical cylinder with an attack angle of $20^{\circ}$}

\begin{figure}
\centering
\includegraphics[width=0.6\textwidth]{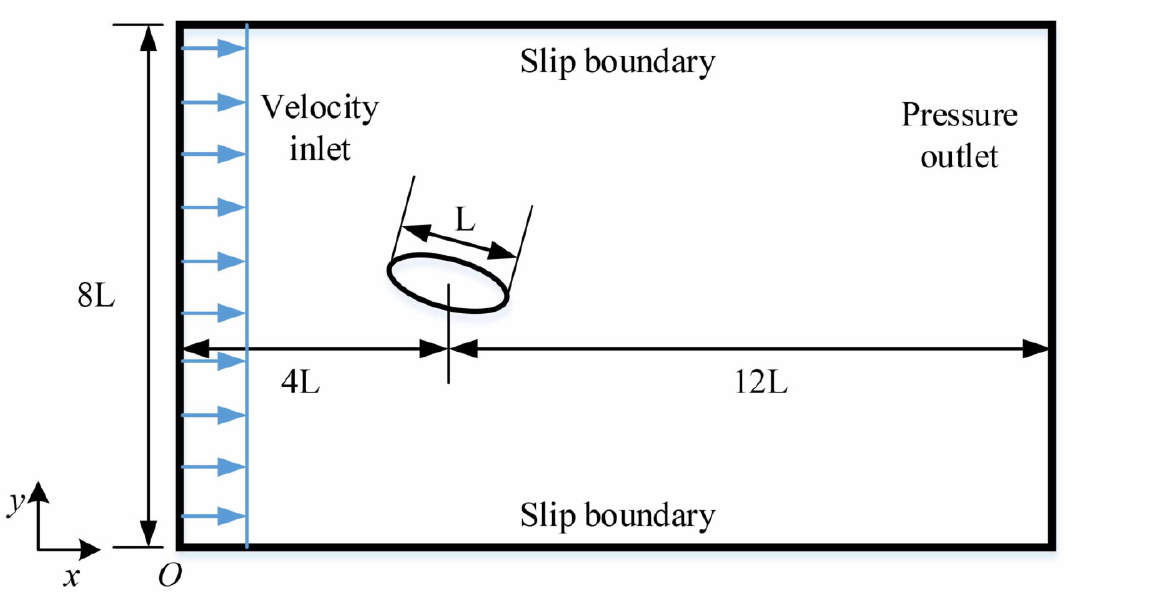}
\caption{Schematic of the flow over an inclined elliptical cylinder with an attack angle of $20^{\circ}$.}
\label{ellip_schm}
\end{figure}

In many scenarios, the vortical structure has high velocity gradient, which is closely related to flow instability, as such the vorticity needs to be tracked and high resolution should be deployed around. The proposed BAPR method is also applicable for detecting and resolving the high vorticity zones. To demonstrate this property, the flow over an elliptical cylinder is simulated with the proposed BAPR method. The setup of this case is shown in Fig. \ref{ellip_schm} \cite{sun2017deltaplus}, where the elliptical cylinder is inclined with an attack angle of $20^{\circ}$, and the major axis length is $L=0.2$ $\rm{m}$ with the axis ratio of 0.4. The fluid domain is $16L$ in width and $8L$ in height. The centroid of the cylinder is positioned at $(4L,4L)$,  {and the refined resolution in this case is set as $\Delta x_{1}=L/60$}. The density of fluid is  $\rho=1000$ $\rm kg/m^{3}$. The flow enters from the left inlet, and the Reynolds number is $Re=UL/\nu=500$, where $\nu=0.001$ $\rm{m^{2}/s}$. For the inlet and outlet boundaries, the boundary conditions are implemented with buffer zones similar to the case in section \ref{sect:Flapbeam}, and a slip boundary condition is deployed at the upper and lower sides.  {The characteristic identification in this case is based on the vorticity $\omega$ with a threshold of $\left|\omega \right|_{th}=30$ $\rm s^{-1}$.}  {The block array sizes $(n_x,n_y)$ in this case are $n_x=20$ and $n_y=10$ in the horizontal and vertical directions, respectively. }

\begin{figure}
\centering
\includegraphics[width=1\textwidth]{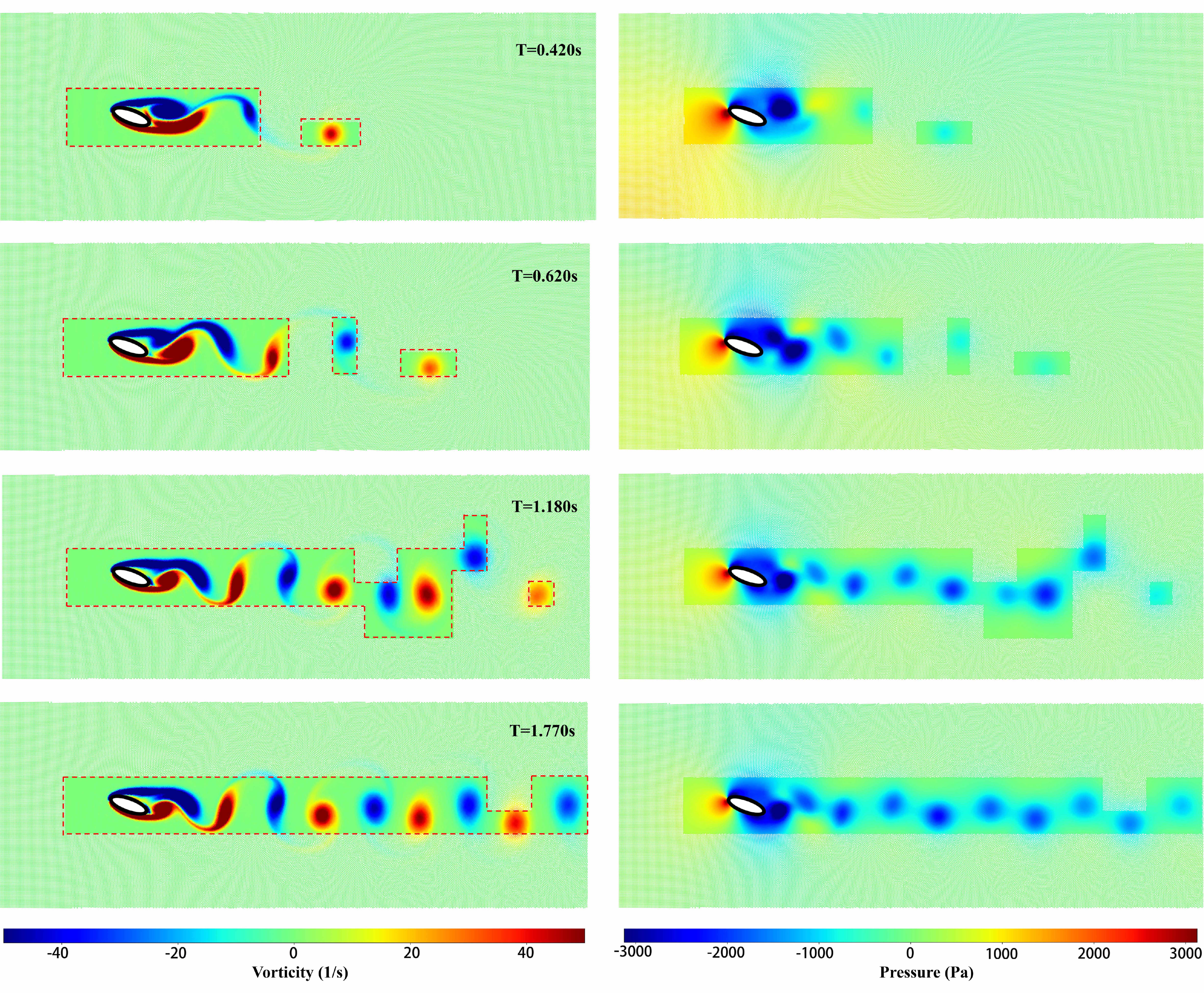}
\caption{Flow over an inclined elliptical cylinder with the particle refinement based on the vorticity identification: the vorticity  (left) and pressure (right) fields. The BAPR refinement zones are shown within the red dashed lines.}
\label{ellip_snap}
\end{figure}

\begin{figure}
\centering

\subfigure{
\begin{minipage}{1\textwidth}
\centering
\includegraphics[width=0.7\textwidth]{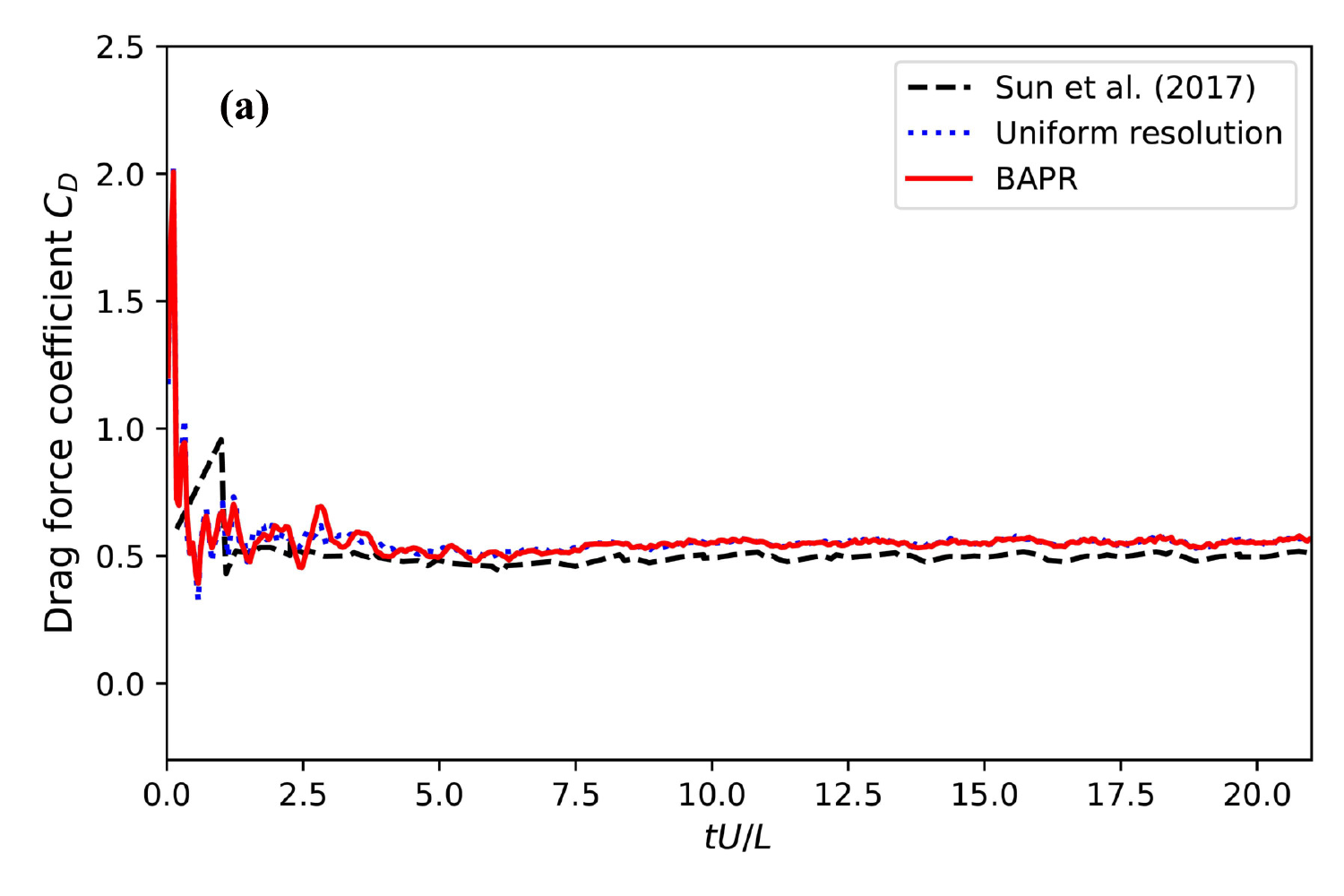}
\end{minipage}
}
\subfigure{
\begin{minipage}{1\textwidth}
\centering
\includegraphics[width=0.7\textwidth]{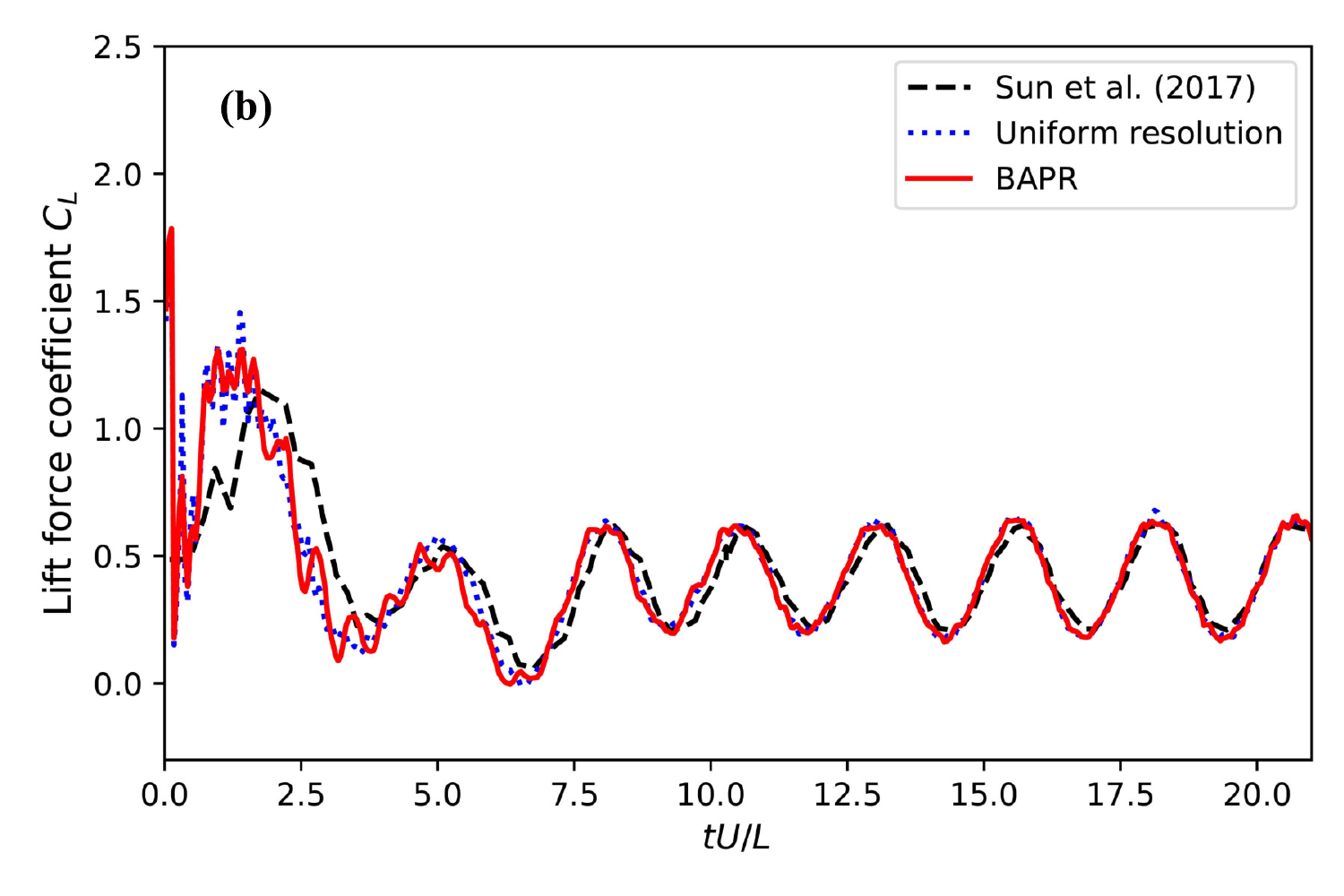}
\end{minipage}
}

\caption{Drag (a) and lift (b) force coefficients of the elliptical cylinder, and the comparisons with the simulation results by Sun et al. \cite{sun2017deltaplus}.}
\label{ellip_plot}
\end{figure}

Fig. \ref{ellip_snap} shows the simulation results with the proposed BAPR method, where the vorticity and pressure fields are given in the left and right columns, respectively. At $t=0.42$ $\rm s$, some regions with vorticity exceeding the threshold are identified and refined, see the regions enclosed with red dashed lines. At $t=0.62$ $\rm s$ and $t=1.18$ $\rm s$, more vortical structures further develop, and the identified blocks are refined adaptively to capture these targeted features. At $t=1.77$ $\rm s$, even more structures are captured, and the refinement blocks can adaptively change the topology for the targeted regions. At the right column, the corresponding pressure fields are shown, and it is illustrated that the pressure fields have a very smooth transition between the refinement and non-refinement zones, due to the deployed regularization technique for newly generated particles. Fig. \ref{ellip_plot} shows the drag and lift force coefficients of the cylinder, and their comparisons with the results from the uniform resolution simulation and the results from \cite{sun2017deltaplus}. It is seen that results from the BARR method and the uniform resolution simulation have a very good agreement with each other, suggesting that the present BAPR method is able to achieve almost the same accuracy as the uniform resolution simulation. While there are some slight oscillations in the present results, the coefficients of $C_{\rm{D}}$ and $C_{\rm{L}}$ both have a reasonable agreement with the references.

The computational time of the BAPR method and the uniform resolution simulation is compared in Table \ref{table:cputime_ellip}. The implementation is based on MPI ranks each containing a set of threads, and apart from the particle resolution, all the settings are the same. As shown in Table \ref{table:cputime_ellip}, when 8 MPI ranks (each involves 1 thread) are employed, for the case with the space resolution $\Delta x_{1}=L/60$, the computational time for each 0.002 $\rm s$ physical simulation time is 181 $\rm s$ with the uniform resolution, while that with BAPR is only 85 $\rm s$. Likewise, for the case with the resolution $\Delta x_{1}=L/80$, the computational time for the uniform resolution simulation and the BAPR method is 499 $\rm s$ and 224 $\rm s$, respectively. When 32 threads are employed without MPI data communication, the computational time of the BAPR method is approximately $1/4$ of that using a uniform resolution, indicating a more remarkable efficiency improvement. It is worth noting that the MPI data communication accounts for a great portion of the total computational time. It is expected that if the MPI communication is optimized at its best, a much higher efficiency can be achieved with the BAPR method. Despite this, it is seen that the BAPR method is still able to save the computational cost significantly, and preserve the solution accuracy at the same time. 

\begin{table}
	\renewcommand{\arraystretch}{1.2}
	\caption{Comparison of the average computational time for simulating each 0.002 $\rm s$ between the uniform resolution simulation and the proposed BAPR method. }
	\label{table:cputime_ellip}
	\begin{center}
		
		\begin{tabular}{p{6cm} p{3cm} p{4cm} }
			\toprule [1.2 pt]
			 Cases &   BAPR & Uniform resolution\\
			\hline
			 $ {\Delta x_1=L/60}$, 8 MPI (1 thread) & 85 $\rm s$ & 181 $\rm s$  \\
			  $ {\Delta x_1=L/80}$, 8 MPI (1 thread) & 224 $\rm s$ & 499 $\rm s$  \\
			 $ {\Delta x_1=L/60}$, 1 MPI (32 threads) & 35 $\rm s$ & 153 $\rm s$  \\
			$ {\Delta x_1=L/80}$, 1 MPI (32 threads)& 94 $\rm s$ & 397 $\rm s$  \\					  
			\bottomrule [1.2 pt]
		\end{tabular}
	\end{center}
\end{table}

\subsection{Body entry problems}
\subsubsection{Single-body water entry problem}
\begin{figure}
\centering
\includegraphics[width=0.6\textwidth]{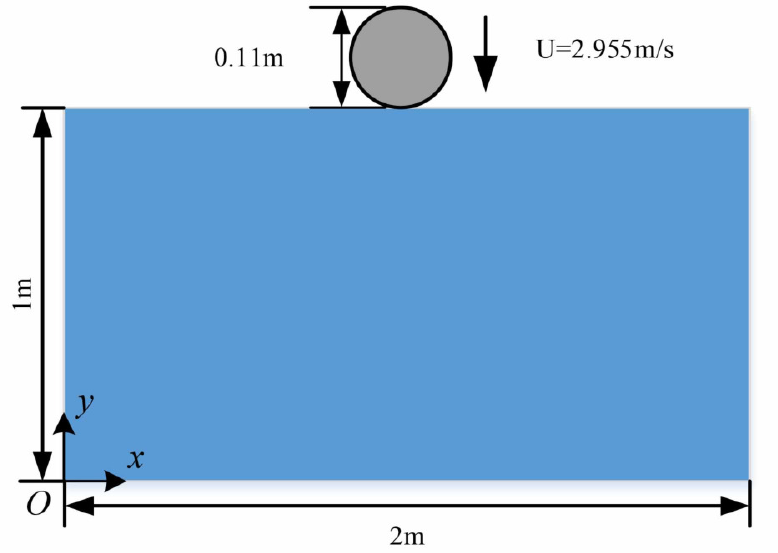}
\caption{Schematic for the water entry of a horizontal cylinder.}
\label{entry_schm}
\end{figure}

In order to further demonstrate the capability of the proposed BAPR method to adaptively track the solid, a single-body water entry problem is investigated in this case. The computational setup is shown in Fig. \ref{entry_schm}, where the water is 2 $\rm m$ in width and 1 $\rm m$ in height, and the diameter of the cylinder is $D=0.11$ $\rm{m}$. The coordinates of the lower end of the cylinder are (1 $\rm m$, 1 $\rm m$). The cylinder enters the water with a velocity of $U=2.955$ $\rm {m/s}$. The density of water and cylinder is 1000 $\rm kg/m^{3}$ and 500 $\rm kg/m^{3}$, respectively.  { The computational resolution is $\Delta x_{1}=D/40$ for the refinement zones.} In this case, the adaptation criterion for the proposed BAPR method is based on the tracking of the solid phase.  {The block array sizes $(n_x,n_y)$ in this case are $n_x=20$ and $n_y=10$ in the horizontal and vertical directions, respectively. } Regarding the formulations for calculating the integrated force and torque on the rigid body, the readers are referred to \cite{bouscasse2013nonlinear}.

\begin{figure}
\centering
\includegraphics[width=0.8\textwidth]{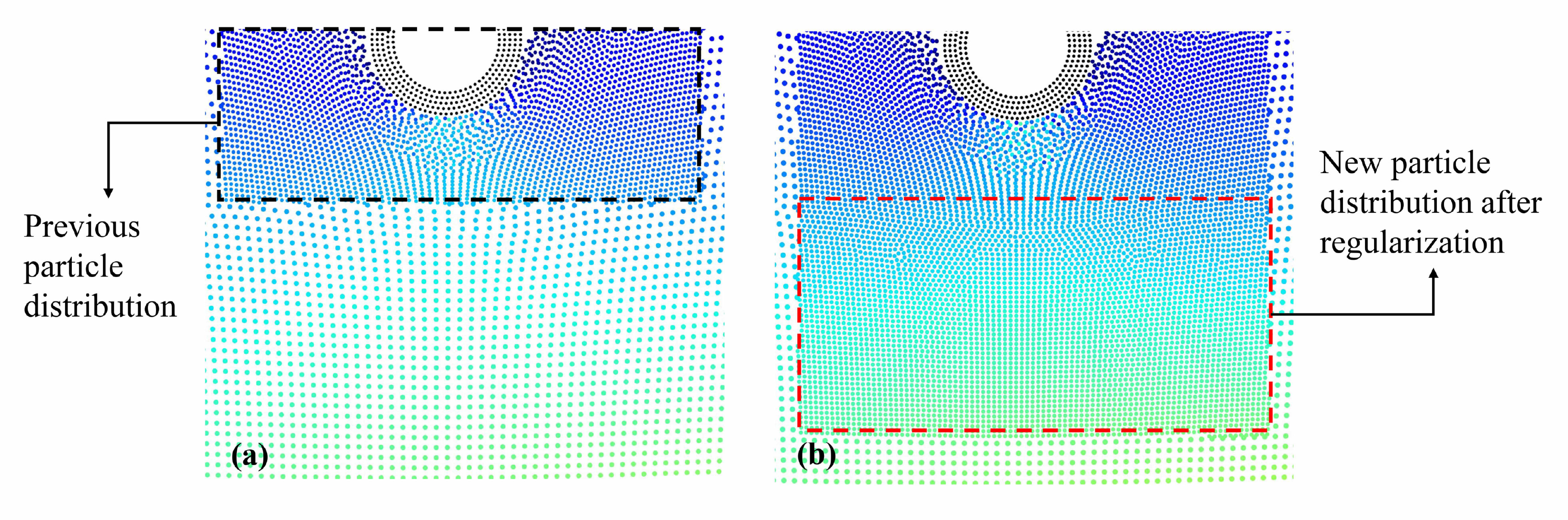}
\caption{Zoomed-in views of the local particle refinement with the present BAPR method :(a) the previous particle distribution; (b) the new particle distribution after deploying the regularization technique.}
\label{entry_link}
\end{figure}

\begin{figure}
\centering
\includegraphics[width=0.95\textwidth]{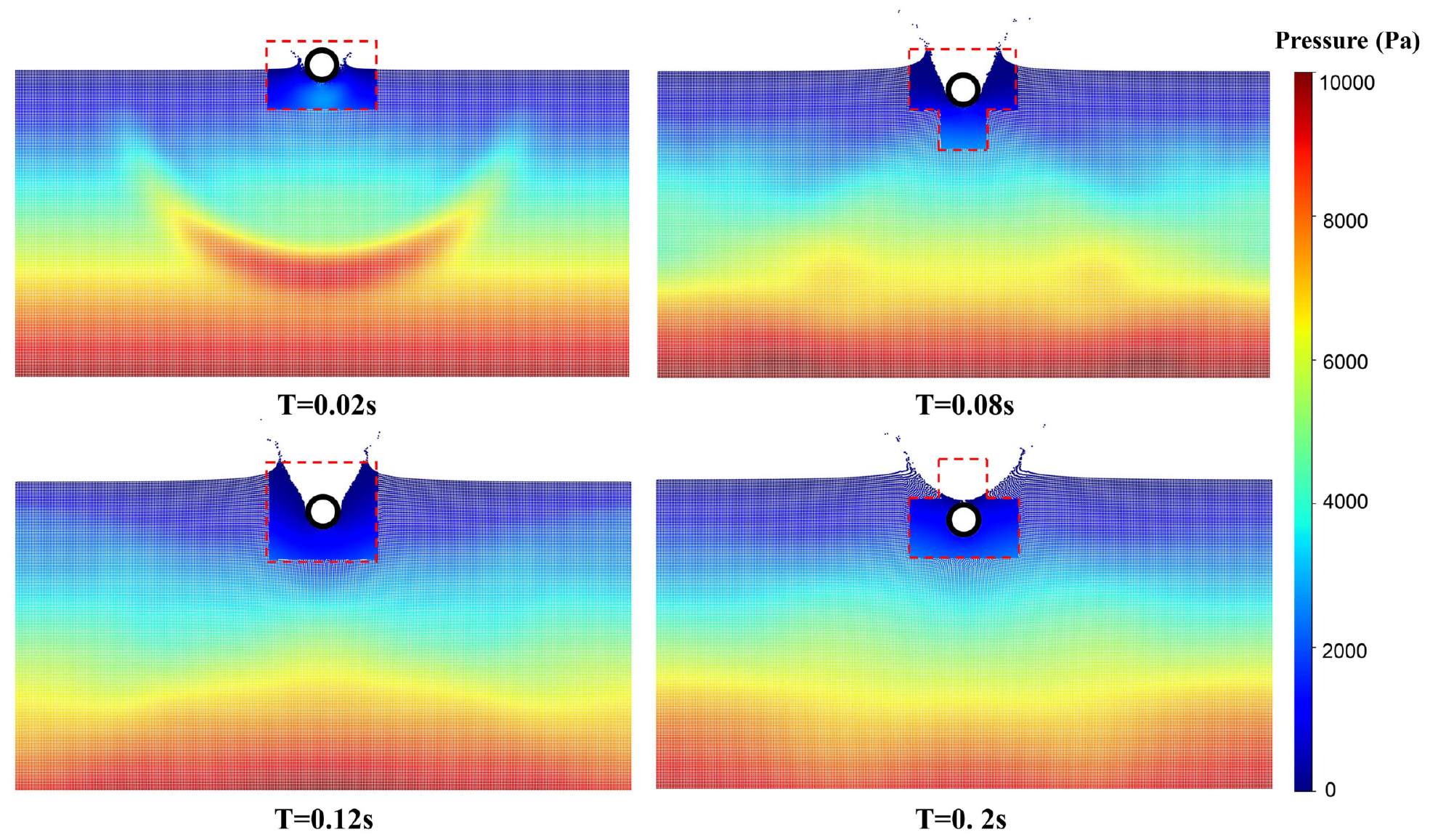}
\caption{Single-body water entry: the simulation results of the pressure field for the water entry of a horizontal cylinder with the BAPR method at different physical time, where the BAPR refinement is based on the characteristic of the `solid' phase.}
\label{entry1}
\end{figure}

Fig. \ref{entry_link} shows the new particles in the region enclosed with the red dashed lines. In Fig. \ref{entry_link} (a), the cylinder goes downwards, thus new block zones are activated with the new particles generated inside. Due to the particle regularization method in section \ref{sect:Regularization}, the particle distribution around the interface between the previous particles and the new particles is very smooth. Fig. \ref{entry1} shows the snapshots of the pressure field simulated with the proposed BAPR method, and the refinement blocks are enclosed with the red dashed lines. The computational domain around the cylinder is always deployed adaptively with the refined particle resolution thanks to the BAPR method. In the entire computational domain, the pressure fields are also very smooth without any spurious oscillations. The entering depth of the cylinder using the present BAPR method (red line) is compared to the simulation results by Yang et al. \cite{yang2021smoothed} and Sun et al. \cite{sun2018accurate}, and the experimental results by Zhu et al. \cite{zhuwater2006}.  It is shown that our results agree very well with those studies.

\begin{figure}
\centering
\includegraphics[width=0.95\textwidth]{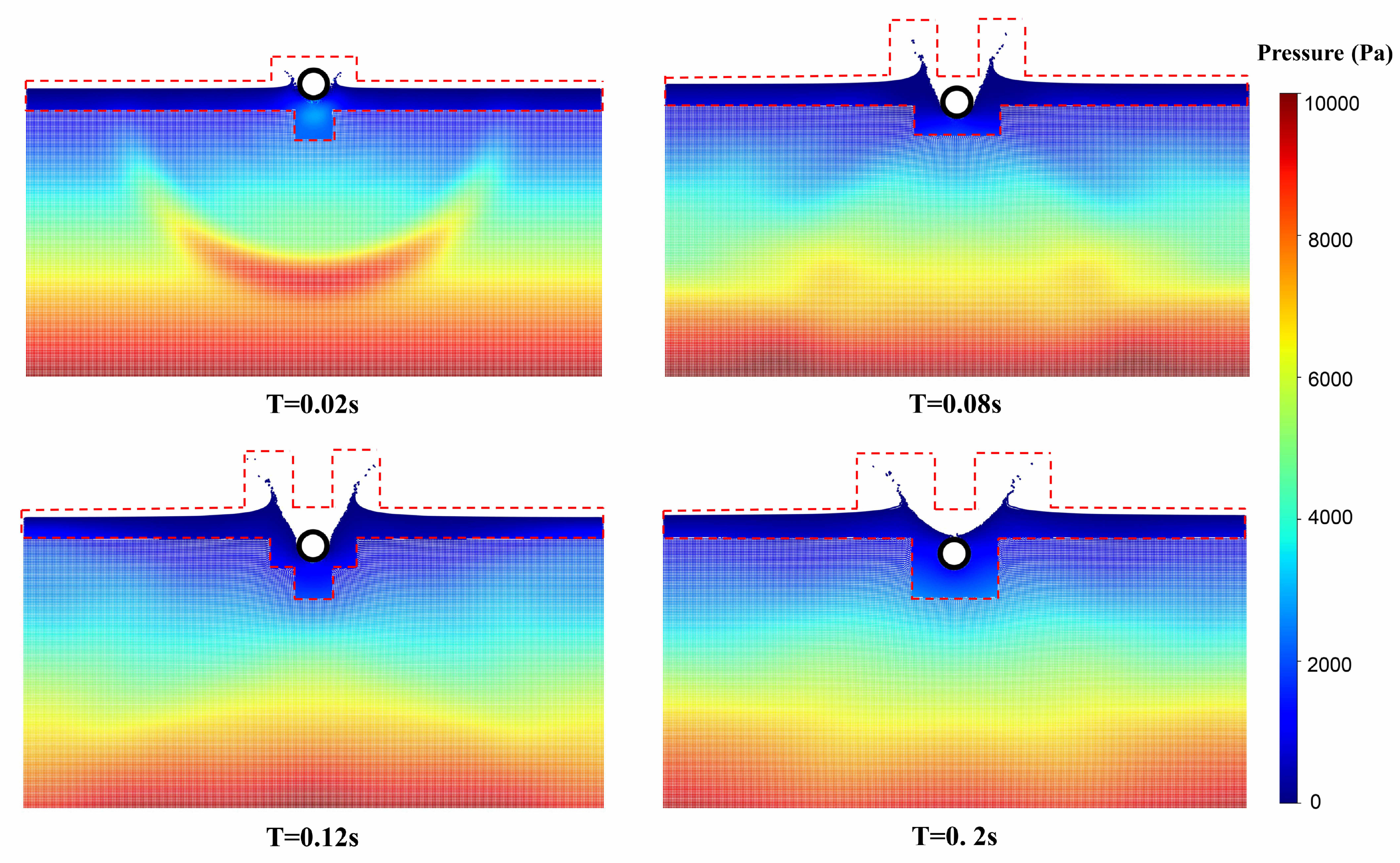}
\caption{ {Single-body water entry: the simulation results of the pressure field for the water entry of a horizontal cylinder with the BAPR method at different physical time, where the BAPR refinement is based on the characteristics of the `solid' phase and the `free surface'.}}
\label{entry12}
\end{figure}

\begin{figure}
\centering
\includegraphics[width=0.7\textwidth]{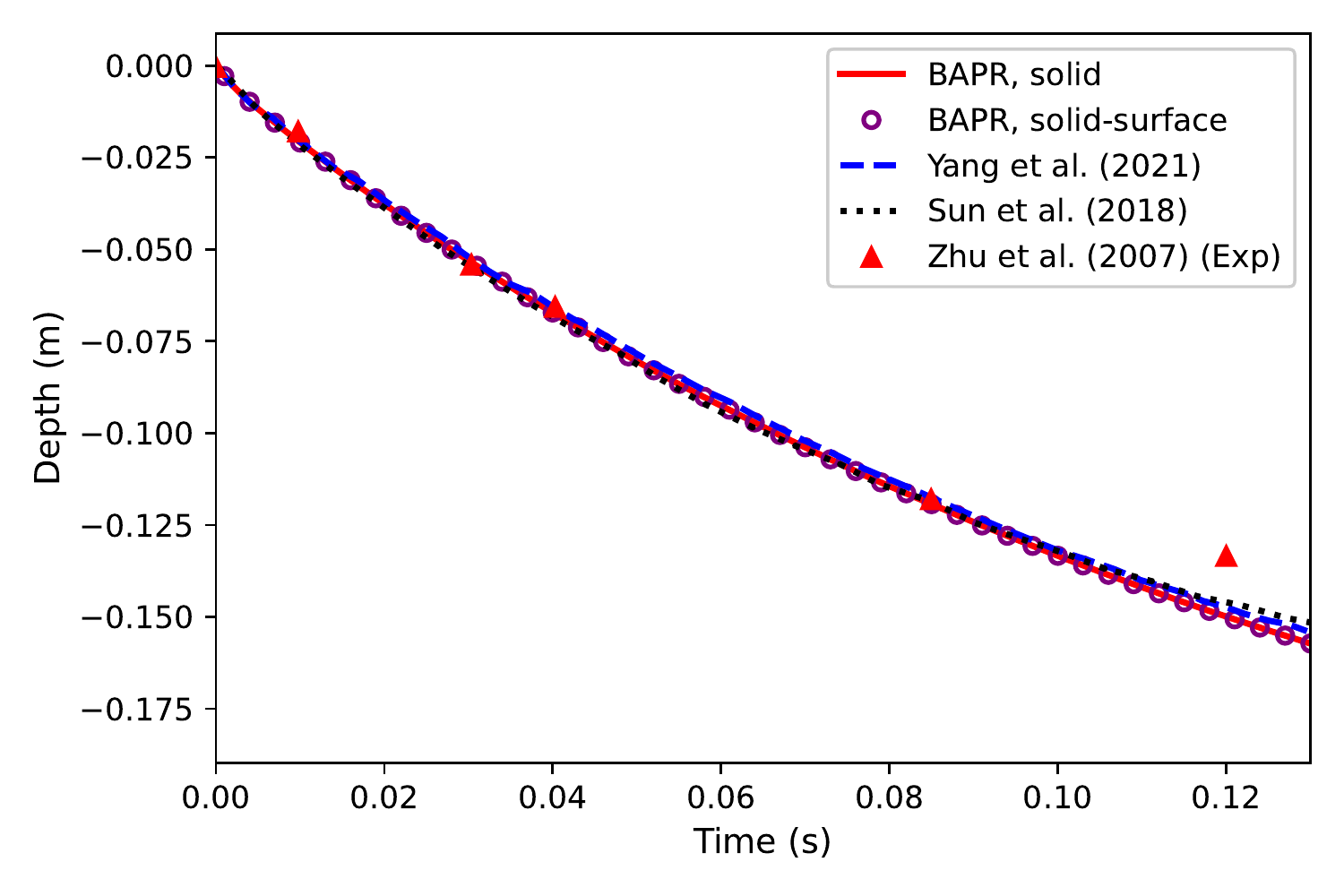}
\caption{ {Depth statistics of the water entry of a horizontal cylinder with BAPR using the characteristics of (1) `solid' (red curve), (2) `solid' and `free surface' (violet circles), and the comparisons with the simulation results of Yang et al. \cite{yang2021smoothed} and Sun et al. \cite{sun2018accurate}, and the experimental data of Zhu et al. \cite{zhuwater2006}.} }
\label{entry_plot}
\end{figure}

 {Sometimes, the free surface region is important and should be solved with a high resolution. In BAPR, the `free surface' can also be selected as a targeted characteristic. Herein, we will demonstrate this with BAPR employing  two targeted characteristics, i.e.,  the `solid' phase and the `free surface'. The `free surface' characteristic can be extracted with the free surface detection method by Marrone et al. \cite{marrone2010fast} before the refinement. For a high block resolution in surface areas, the block array sizes $(n_x,n_y)$ are chosen as $n_x=24$ and $n_y=12$ in the horizontal and vertical directions, respectively, in this simulation. 
Fig. \ref{entry12} displays the snapshots of water entry with BAPR using the characteristics of `solid' phase and `free surface'. It is shown that both the surface region and the solid region are refined adaptively, and the entry depth (indicated by violet circles in Fig.~\ref{entry_plot}) agrees very well with the simulation using the characteristic of `solid' phase. }

\subsubsection{Multi-body water entry problem}

\begin{figure}
\centering
\includegraphics[width=0.6\textwidth]{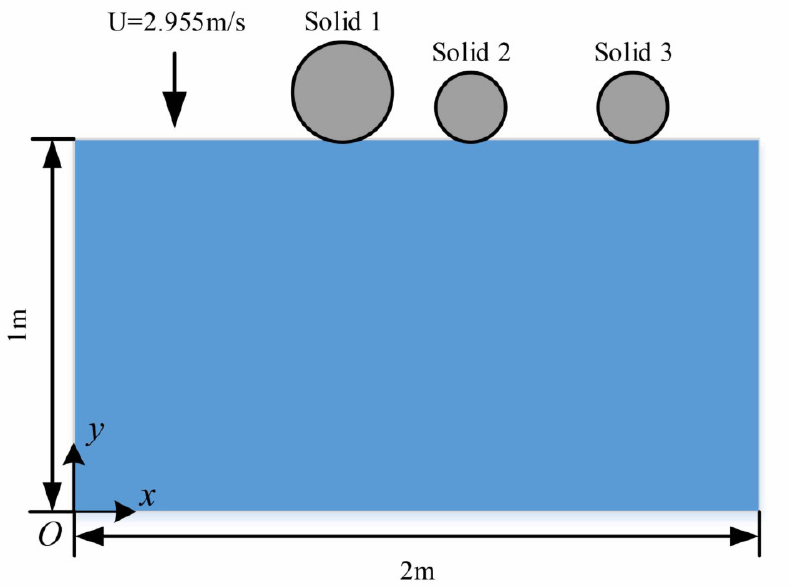}
\caption{Schematic of the multi-body water entry problem.}
\label{entry_M_schm}
\end{figure}

In this case, we consider the multi-body water entry problem, with which the proposed BAPR method will show its potential for complex multi-body problems. The corresponding setup is shown in Fig. \ref{entry_M_schm}, i.e., three cylinders fall into the water domain with an initial velocity of $U=2.955$ $\rm{m/s}$. The coordinates of the lower ends of the three cylinders are (0.8 $\rm m$, 1 $\rm m$), (1.05 $\rm m$, 1 $\rm m$) and (1.45 $\rm m$, 1 $\rm m$); their density is $500$ $\rm{kg/m^{3}}$, $500$ $\rm{kg/m^{3}}$ and $2000$ $\rm{kg/m^{3}}$; their diameter is $D_{1}=0.11$  $\rm{m}$, $D_{2}=0.08$ $\rm{m}$ and $D_{3}=0.08$ $\rm{m}$, respectively. The kinetic viscosity of the fluid is set as $ 0.001$ $\rm{m^2/s}$ to avoid turbulence around 
Solid 3.  {The refined resolution for the solid phase is $\Delta x_{1}=D_{1}/40$. }  {The block array sizes $(n_x,n_y)$ in this case are $n_x=20$ and $n_y=10$ in the horizontal and vertical directions, respectively. }

\begin{figure}
\centering
\includegraphics[width=0.9\textwidth]{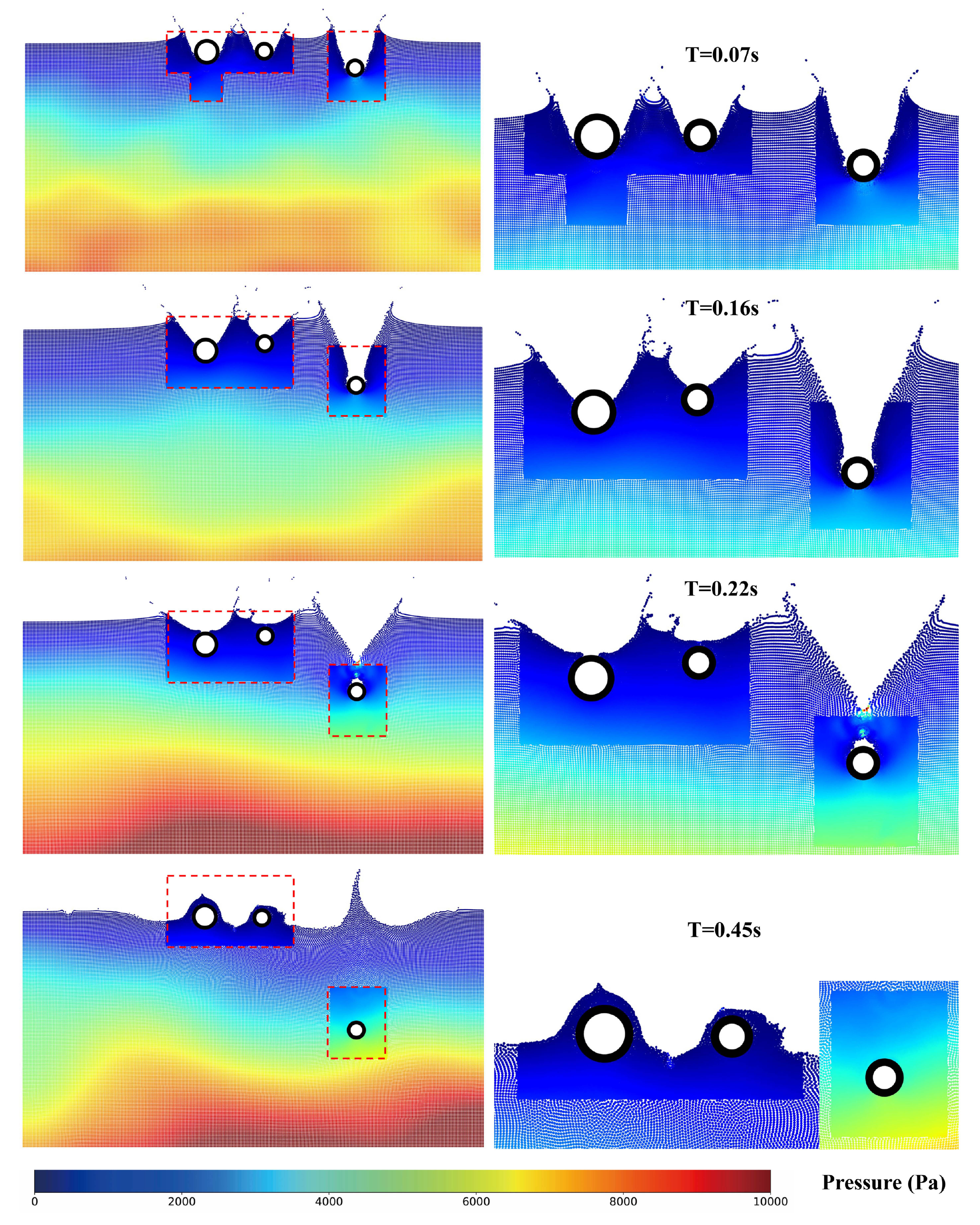}
\caption{The pressure fields of multi-body water entry at different physical time (left) and the zoomed-in details (right).}
\label{entry_M_snap}
\end{figure}

\begin{figure}
\centering
\includegraphics[width=0.7\textwidth]{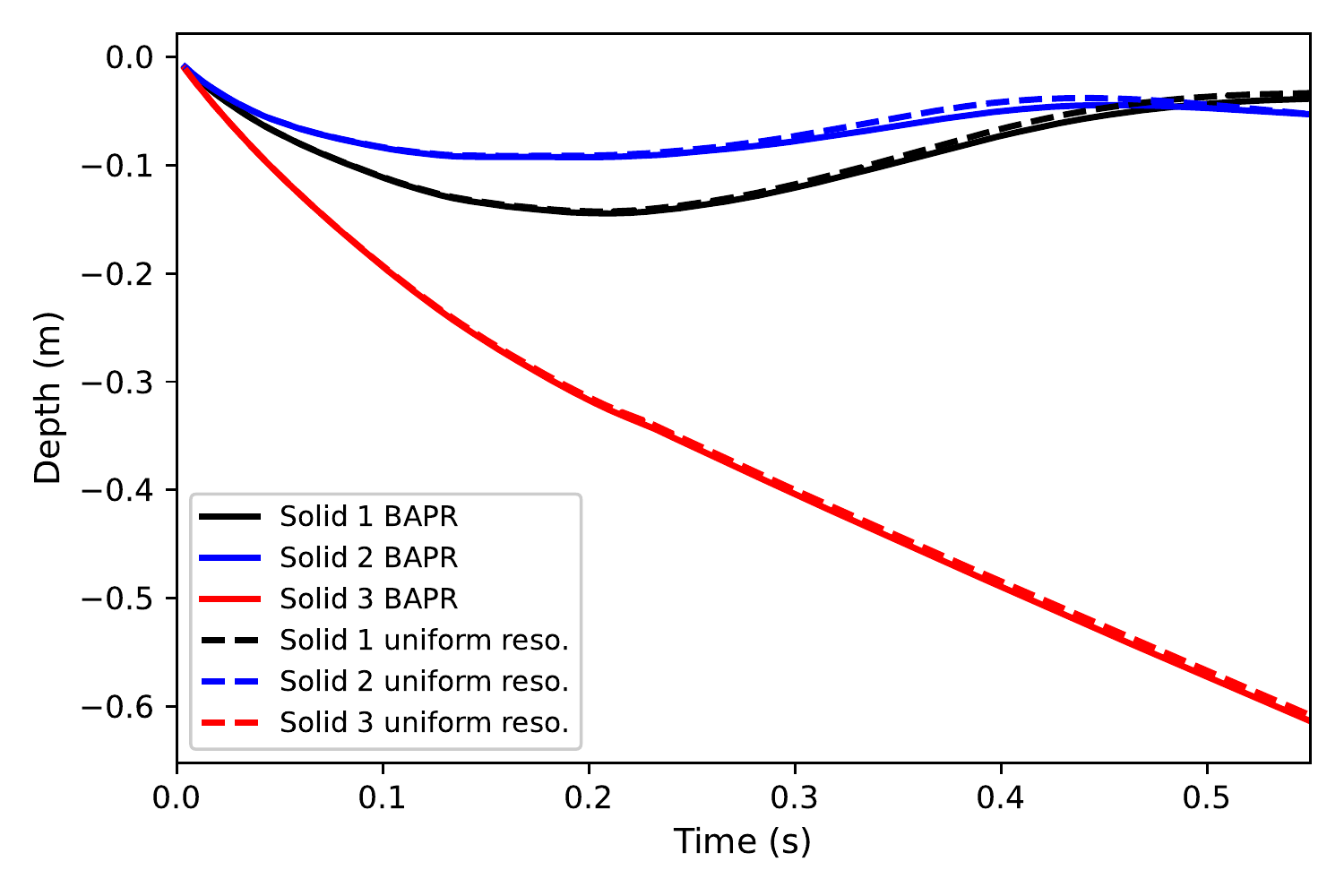}
\caption{Depth statistics of multi-body water entry using the BAPR method and the comparisons with those from the uniform resolution simulation.}
\label{entry_M_plt}
\end{figure}

The snapshots of the multi-body water entry simulation are shown in Fig. \ref{entry_M_snap}. The refinement blocks are enclosed with red dashed lines. It is shown that the three cylinders are always surrounded by the refinement blocks with the BAPR method throughout the process. For Solid 1 and Solid 2, the refinement blocks are activated and linked with each other because the two solids are too close. For Solid 3, it sinks very fast as its density is larger than other two solids, and it is shown that the surrounding block zones are activated adaptively throughout the falling process. It is worth noting that the pressure field is very smooth throughout the domain, owing to the particle regularization at the initialization stage. In order to validate the accuracy of the BAPR method, the uniform resolution simulation is also conducted with the identical setup. The entering depths of the three solids are shown in Fig. \ref{entry_M_plt}. It is shown that the results from the present BAPR method almost overlap with that of the uniform resolution simulation. The computational time deploying different MPI ranks and threads for this problem is compared between the uniform resolution simulation and the BAPR method in Table \ref{table:cputime_entry1}, which shows that the BAPR method can have an approximately larger than $50\%$ efficiency improvement. To the best of authors' knowledge, this is the first SPH study on the multi-body water entry problem with an adaptive multi-resolution method. This case illustrates the potential of the proposed BAPR method for simulating the complex problems in the ocean engineering. 

\begin{table}
	
	\renewcommand{\arraystretch}{1.2}
	\caption{Comparison of the average computational time for simulating each 0.002 $\rm s$ between the uniform resolution simulation and the proposed BAPR method. }
	\label{table:cputime_entry1}
	\begin{center}
		\begin{tabular}{p{6.5cm} p{2cm} p{4cm} }
			\toprule [1.2 pt]
			Cases &   BAPR & Uniform resolution\\
			\hline
			$ {\Delta x_1=D_1/40}$, 4 MPI (1 thread) & 86 $\rm s$ & 173 $\rm s$  \\
			$ {\Delta x_1=D_1/60}$, 4 MPI (1 thread) & 294 $\rm s$ & 590 $\rm s$  \\
			$ {\Delta x_1=D_1/40}$, 1 MPI (32 threads) & 14 $\rm s$ & 35 $\rm s$  \\
			$ {\Delta x_1=D_1/60}$, 1 MPI (32 threads) & 42 $\rm s$ & 90 $\rm s$  \\				
			\bottomrule [1.2 pt]
		\end{tabular}
	\end{center}
\end{table}

\section{Conclusions}
\label{section:5}

In this study, a BAPR method is proposed to refine the targeted simulation zones adaptively based on the characteristic identification. In this method, the block array is deployed and the activation status of the transition and activation zones are determined with novel algorithms. A particle regularization technique is proposed for a more isotropic distribution of the newly generated particles in the newly activated blocks. The numerical schemes for solving FSI problems are developed with the present BAPR method. The framework incorporates the ALE formulations for the fluid phase and the total Lagrangian formulations for the solid phase. Various benchmark cases, including water impact on an elastic beam, dam-breaking through an elastic beam, flapping beam induced by the flow over a cylinder, flow over an inclined elliptical cylinder, and body entry problems, have been simulated with the BAPR method. Numerical results demonstrate that the present BAPR method shows many advantages, as follows.

(a) The simulation results from the present BAPR method have almost the same accuracy with the uniform resolution simulation;

(b) The particle refinement of the BAPR method can be based on any targeted characteristic in the computational field, e.g., phase, vortex, velocity, etc.;

(c) The BAPR method can track the targeted characteristic adaptively;

(d) The BAPR method can save the computational time significantly compared to the uniform resolution simulation.

The presented numerical framework is 2D with two resolution levels.   {The extension of the present framework to 3D with more than two resolution levels will be investigated in the future work}. In a word, the present BAPR framework provides an efficient and accurate multi-resolution method for solving complex fluids and FSI problems.

\section*{Declaration of Competing Interest}

The authors declare that they have no known competing financial interests or personal relationships that could have appeared to influence the work reported in this paper.

\section*{Data availability}
The data that support the findings of this study are available on request from the corresponding author, Lin Fu.

\section*{Acknowledgment}

This work was supported by the Research Grants Council (RGC) of the Government of Hong Kong Special Administrative Region (HKSAR) with RGC/GRF Project (No. 16206321) and the fund from Shenzhen Municipal Central Government Guides Local Science and Technology Development Special Funds Funded Projects (No. 2021Szvup138). Lin Fu also acknowledges Zhe Ji, who has been working with him to develop the basic SPH platform over the past several years.

  \bibliographystyle{elsarticle-num}
  \scriptsize
  \setlength{\bibsep}{0.5ex}

\bibliography{bib.bib}

\end{document}